%% file: sample-sigconf.tex
\documentclass[sigconf]{acmart}
\usepackage{graphicx}
\usepackage{ulem}
\usepackage{verbatim}
\usepackage{multirow}
\usepackage[linesnumbered, ruled, vlined]{algorithm2e}
\usepackage{longtable}
\usepackage{array}
\usepackage{subfigure}
\usepackage{stfloats}
\usepackage{balance}
\usepackage{multicol}
\usepackage{color}
\usepackage{bm}
\usepackage{booktabs} % For formal tables
\usepackage{epsfig}
\usepackage{enumitem}
\usepackage{balance}
\usepackage{makecell}

\usepackage{arydshln}

\usepackage{cleveref}
\usepackage{graphicx}
\usepackage{textcomp}
\usepackage{xcolor}
\usepackage{subfigure}

\usepackage{tabularx}

\usepackage{makecell}

\newcommand{\hide}[1]{} %hide
\newcommand{\vpara}[1]{\vspace{0.1in}\noindent\textbf{#1}}

\newcommand{\secref}[1]{Section~\ref{#1}} %section reference
\newcommand{\beqn}[1]{\vspace{-0.03in}\begin{eqnarray}#1\end{eqnarray}\vspace{-0.03in}}

\newcommand{\RC}{MOP}
\newcommand{\sRC}{MOP\space}
\AtBeginDocument{%
  }

\setcopyright{acmcopyright}
\copyrightyear{2018}
\acmYear{2018}
\acmDOI{XXXXXXX.XXXXXXX}

\acmConference[Conference 'XX]{XX}{June XX,
  2018}{NY}

\acmPrice{15.00}
\acmISBN{978-1-4503-XXXX-X/18/06}

\begin{document}

%%
%% The "title" command has an optional parameter,
%% allowing the author to define a "short title" to be used in page headers.
\title{Motif-Based Prompt Learning for Universal Cross-Domain Recommendation}

%%
%% The "author" command and its associated commands are used to define
%% the authors and their affiliations.
%% Of note is the shared affiliation of the first two authors, and the
%% "authornote" and "authornotemark" commands
%% used to denote shared contribution to the research.
\author{Bowen Hao}
\email{6974@cnu.edu.cn}
\orcid{1234-5678-9012}
\affiliation{%
  \institution{Capital Normal University}
  \city{Beijing}
  \country{China}
}

\author{Chaoqun Yang}
\email{chaoqun.yang@griffith.edu.au}
\affiliation{%
	\institution{Griffith University}
	\city{Gold Coast}
	\country{Australia}
}

\author{Lei Guo}
\email{leiguo.cs@gmail.com}
\affiliation{%
	\institution{Shandong Normal University}
	\city{JiNan}
	\country{China}
}

\author{Junliang Yu}
\email{ jl.yu@uq.edu.au}
\affiliation{%
	\institution{The University of Queensland}
	\city{Brisbane}
	\country{Australia}
}

\author{Hongzhi Yin}
\email{h.yin1@uq.edu.au}
\authornote{Corresponding Author}
\affiliation{%
  \institution{The University of Queensland}
  \city{Brisbane}
  \country{Australia}}

%%
%% By default, the full list of authors will be used in the page
%% headers. Often, this list is too long, and will overlap
%% other information printed in the page headers. This command allows
%% the author to define a more concise list
%% of authors' names for this purpose.
\renewcommand{\shortauthors}{XX et al.}

%%
%% The abstract is a short summary of the work to be presented in the
%% article.
\begin{abstract}
Cross-Domain Recommendation (CDR) stands as a pivotal technology addressing issues of data sparsity and cold start by transferring general knowledge from the source to the target domain. However, existing CDR models suffer limitations in adaptability across various scenarios due to their inherent complexity.
To tackle this challenge, recent advancements introduce universal CDR models that leverage \textit{shared embeddings} to capture general knowledge across domains and transfer it through ``Multi-task Learning''  or ``Pre-train, Fine-tune'' paradigms.
However, these models often overlook the broader structural topology that spans domains and fail to align training objectives, potentially leading to negative transfer.
To address these issues, we propose a motif-based prompt learning framework, \RC, which introduces \textit{motif-based shared embeddings} to encapsulate generalized domain knowledge, catering to both intra-domain and inter-domain CDR tasks. 
Specifically, we devise three typical motifs: butterfly, triangle, and random walk, and encode them through a Motif-based Encoder to obtain motif-based shared embeddings.
Moreover, we train \sRC under the ``Pre-training \& Prompt Tuning''  paradigm. By unifying pre-training and recommendation tasks as a common motif-based similarity learning task and integrating adaptable prompt parameters to guide the model in downstream recommendation tasks, \sRC excels in transferring domain knowledge effectively.
Experimental results on four distinct CDR tasks demonstrate the effectiveness of \sRC than the state-of-the-art models.
\end{abstract}

%%
%% The code below is generated by the tool at http://dl.acm.org/ccs.cfm.
%% Please copy and paste the code instead of the example below.
%%



%%
%% Keywords. The author(s) should pick words that accurately describe
%% the work being presented. Separate the keywords with commas.
%\keywords{Cross-domain Recommendation, Motif, Prompt Learning}
%% A "teaser" image appears between the author and affiliation
%% information and the body of the document, and typically spans the
%% page.

%%
%% This command processes the author and affiliation and title
%% information and builds the first part of the formatted document.
\maketitle

\input{intro}
\input{related}

\input{approach}

\input{exp}

\input{conclusion}
\input{ack}

\normalem %add this .
\bibliographystyle{ACM-Reference-Format}
\bibliography{sample-base}

\end{document}

%% file: intro.tex
\section{Introduction}

Recommendation systems (RS) play a crucial role in discovering users' latent interests and preferences, thereby being embraced by numerous E-commerce platforms and content providers for driving incremental revenue~\cite{conf/icde/YinW0LYZ19,conf/kdd/YuY000H21}. However, existing RS still suffer from data sparsity and cold-start problems as most users only interact with a fraction of available items, and newly onboard users exhibit no initial interactions. In response to these hurdles, cross-domain recommendation (CDR) is proposed to transfer knowledge from domains with abundant user-item interactions (source) to the domain with few interactions (target). This endeavor seeks to enhance the recommendation performance within the target domain via shared users/items.

\begin{figure}[t]
	\centering
	\includegraphics[width= 0.45 \textwidth]{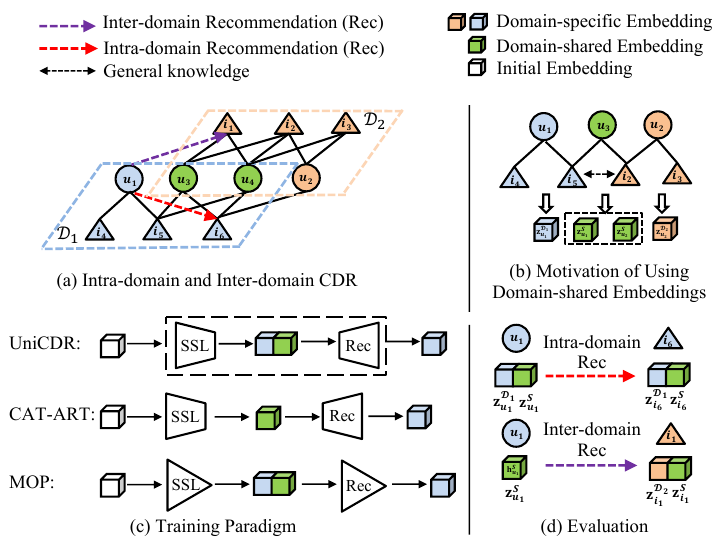}
	\caption{\label{fig:introduction} The technical route of universal CDR model. }
 \vspace{-20pt}
\end{figure}

In general, CDR tasks can be divided into two distinct categories: intra-domain and inter-domain CDR. Intra-domain CDR operates by harnessing knowledge from other domains to enhance recommendation performance for users who possess limited interactions within the current domain (e.g., recommending item $i_6$ within the domain to user $u_1$, as highlighted by the red dotted line in Fig.~\ref{fig:introduction} (a)). Existing research in this line devises transfer modules such as dual~\cite{conf/cikm/HuZY18} or Bi-directional~\cite{conf/cikm/LiuLLP20} neural networks upon the overlapped users/items to achieve knowledge transfer. Inter-domain CDR caters to cold-start users in the new scenario (e.g., recommending $i_1$ for $u_1$, as shown by the purple dotted line in Fig.~\ref{fig:introduction} (a)). This task is usually more challenging as cold-start users have no interactions in the new scenario. To address this issue, researchers focus on designing mapping  functions~\cite{conf/wsdm/ZhuTLZXZLH22,conf/cikm/KangHLY19,conf/sigir/ZhuGZXXZL021}  upon the overlapped users to build the bridge across domains. However, previous endeavors to address these two types of tasks have often necessitated the design of distinct modules for each scenario, demanding substantial effort and restricting the seamless application of these models to downstream tasks.

Recently, researchers have proposed several universal CDR models to support both intra-domain and inter-domain CDR tasks.
These models adhere to a common training paradigm that unfolds as follows: 1) They first integrate users' common interests into \textit{shared embeddings} to capture the general knowledge across domains (as shown in Fig.~\ref{fig:introduction} (b)), with which they can apply their models to both intra-domain and inter-domain CDR tasks.
% The core idea behind these models is that:
% 1) They use \textit{shared embeddings} to capture the general knowledge across domains in a semantic way, which can support both intra-domain and inter-domain CDR tasks.
% As shown in Figure~\ref{fig:introduction}b), the overlapped user $u_3$  interacts with items $i_2$  and $i_5$  from different domains, indicating that $i_2$  and $i_5$  possess shared general knowledge. Since $u_1$  interacts with $i_2$, we can assign domain-shared embeddings $\textbf{z}_{u_1}^{S}$ to $u_1$, which can be used for both intra-domain and inter-domain CDR tasks (Cf. Figure~\ref{fig:introduction}d)).
2) Then, to transfer the general knowledge to the target domain, the ``Multi-task Learning''~\cite{conf/wsdm/CaoL0GL023} (MTL) or ``Pre-train, Fine-tune''~\cite{conf/wsdm/LiXYHLSQN23} (PF) training strategies are often applied. For example, CAT-ART ~\cite{conf/wsdm/LiXYHLSQN23} first pre-trains the model using 
self-supervised learning (SSL) technique such as contrastive learning (CL) and embedding reconstruction (ER) to obtain domain-shared embeddings~\cite{journals/corr/abs-2203-15876}. These embeddings are then transferred to domain-specific embeddings in the fine-tuning stage. UniCDR~\cite{conf/wsdm/CaoL0GL023} adopts MTL, which simultaneously performs CL and recommendation tasks to learn both domain-specific and domain-shared embeddings. Their training paradigms are elucidated in Fig.~\ref{fig:introduction} (c).
% Second, these models transfer the general knowledge into the target domain through several training paradigms such as ``Multi-task Learning''~\cite{conf/wsdm/CaoL0GL023} (MTL) and ``Pre-train, Fine-tune''~\cite{conf/wsdm/LiXYHLSQN23} (PF).
% For example,  as shown in Figure~\ref{fig:introduction}c), 
% CAT-ART~\cite{conf/wsdm/LiXYHLSQN23} adopts PF, which first pre-trains the model using self-supervised learning (SSL) technique such as contrastive learning (CL) and embedding reconstruction (ER) to obtain domain-shared embeddings. These embeddings are then transferred to domain-specific  embeddings in the fine-tuning stage.
% UniCDR~\cite{conf/wsdm/CaoL0GL023} adopts MTL, which simultaneously performs the CL and Rec tasks to learn both domain-specific and domain-shared embeddings.
%图1中的SSL跟CL不一致
Nevertheless, both MTL and PF are not the optimal solutions to the challenges of universal CDR due to the divergence in training objectives. As the pre-training and recommendation tasks are not aligned, the obtained domain-shared embeddings hence may encapsulate spurious correlations, which are less effective in enhancing recommendation performance. Moreover, both approaches ignore that the general structural topology across domains is an essential type of transferable knowledge. For example, as shown in Fig.~\ref{fig:introduction} (a), both the local topology $\mathcal{B}_1$ = \{$u_3$, $i_5$, $u_4$, $i_6$\} in domain $\mathcal{D}_1$ and $\mathcal{B}_2$ = \{$i_2$, $u_4$, $i_3$, $u_2$ \} in domain $\mathcal{D}_2$ reflect similar user preferences that can be transferred across domains. 
% Although the recent proposed ``Pre-train, Prompt, Fine-tune'' (PPT) paradigm can avoid negative transfer by unifying both the SSL and downstream Rec tasks into a common template task such as calculating sub(graph) similarity~\cite{conf/www/LiuY0023} and token pair similarity~\cite{conf/kdd/SunZHWW22}, this training paradigm has not yet been applied to the universal CDR models.

To overcome the above challenges, we resort to the ``Pre-training \& Prompt Tuning'' paradigm~\cite{conf/kdd/SunZHWW22,conf/www/LiuY0023} which enables us to unify pre-training and downstream recommendation tasks into a common template task to address the mismatch issue between the training objectives of pre-training and recommendation, and introduce \textit{motif}~\cite{milo2002network} to capture the general structural topology across domains. 
We name our proposed model as \RC, which is short for \underline{Mo}tif-based \underline{P}rompt Learning for Universal CDR.
%To overcome the above challenges, we resort to the ``Pre-train, Prompt, Fine-tune'' (PPT) paradigm, which enables us to unify both the SSL and downstream Rec tasks into a common template task, such as calculating sub (graph) similarity~\cite{conf/www/LiuY0023} and token pair similarity~\cite{conf/kdd/SunZHWW22}, and propose a Motif-based Prompt Learning Model for Universal CDR.
%先说一下整体的过程吧
% In this work, we propose MOP, a Motif-based Prompt Learning Model for Universal CDR. 
% Unlike existing Universal CDR Models, MOP introduces \textit{motif}~\cite{milo2002network} to capture the general knowledge across domains that can be applied to both intra-domain and inter-domain CDR tasks. Additionally, \sRC leverages  PPT  to effectively transfer general knowledge into the target domain.
% As discussed before, previous works assign each user/item with a semantic-based domain-shared embeddings to capture the general knowledge. However, they 
% ignore the fact that the general structural topology across domains is also a type of general knowledge.
% For example, 
% as shown in Figure~\ref{fig:introduction}a), both the local topology $\mathcal{B}_1$ = \{$u_3$, $i_5$, $u_4$, $i_6$\} in domain $\mathcal{D}_1$ and $\mathcal{B}_2$ = \{$i_2$, $u_4$, $i_3$, $u_2$ \} in domain $\mathcal{D}_2$ reflect the relationship that \textit{Cloth friends always have similar tastes towards items}.
%这个过程是不是分两步，1是学shared embedding, 然后微调
%我觉得下面写的太细节了，轮廓反而不清楚
Specifically, we assign each user/item \textit{motif-based domain-specific embeddings} (M-specific) that contain the structural topology within domains and further introduce \textit{motif-based domain-shared embeddings} (M-shared) to capture the general structural topology across domains, because motif~\cite{milo2002network} involves the structural correlation of nodes. To obtain M-specific and M-shared embeddings, we propose a Motif-based Encoder consisting of a hypergraph-based motif encoding module, a novel Mixture-of-Domain-Experts (MoDE) transformer and an adaptable \textsc{ReadOut} function. Intuitively, the hypergraph can capture complex relationships among nodes within a motif, while the MoDE transformer can handle both M-shared and M-specific embeddings through several  MoDE experts.
Fig.~\ref{subfig:overview1} shows how the Motif-based Encoder works.
%Concretely, we first feed the sampled motifs into the hypergraph module to generate M-specific and M-shared embeddings. These embeddings are then fed into the MoDE Transformer, which routes the M-shared or M-specific experts for further refinement. Finally, we employ a \textsc{ReadOut} function on the refined M-specific and M-shared embeddings  to obtain the central user/item embeddings that contain general structural topology information.
%Since nodes from different domains exhibit both long-term and short-term dependencies~\cite{journals/tois/HaoYZLC23}, we introduce three typical motifs~\cite{milo2002network} that can effectively represent the connection patterns of nodes in the user-item graph: random walk~\cite{conf/sigir/CaoCLW22}, butterfly~\cite{conf/kdd/Sanei-MehriST18} and triangle~\cite{conf/www/YuYLWH021}, for capturing long-term dependencies of nodes, short-term dependencies of nodes with strong connections, and  short-term dependencies of nodes with weak connections, respectively.
% \hide{
% 	Both the local topology $\mathcal{T}_1$  = \{$u_1$, $i_5$, $u_3$\} in domain $\mathcal{D}_1$  and $\mathcal{T}_2$  = \{ $u_2$, $i_2$, $u_4$\} in domain $\mathcal{D}_2$ relect the relationships that \textit{Users with sparse relation may have the same taste towards items}. 
% 	Both the local topology $\mathcal{R}_1$=\{$u_1$,$i_5$,$u_3$,$i_6$,$u_4$\} and $\mathcal{R}_2$ = \{ $u_3$,$i_1$,$u_3$,$i_2$,$u_2$ \} can be viewed as reformatting the  graph data into the structral sequence to capture the corrlations between e
% } 
To seamlessly and effectively transfer the general topology knowledge into the target domain, we propose a prompt tuning strategy, where we first transform the SSL and recommendation tasks in our framework into motif-based similarity learning (MSL) tasks, which calculate the cosine similarity between motif-based node embeddings. Then we pre-train \sRC using SSL tasks to obtain M-specific and M-shared embeddings. Finally, during the prompt tuning stage, in each target domain, we add extra learnable parameters upon the $\textsc{ReadOut}$ function to transfer the M-specific and M-shared embeddings into refined domain-specific
embeddings for performing recommendation task (Cf. Fig.~\ref{fig:overview} (c)).

\hide{Each positive pair in both the CL and ER tasks are the same node with different augmentations, while each negative pair in these two tasks are different nodes with different augmentations.
	The positive and negative pairs in the Recommendation task are the corresponding <user, item> interaction records.}

In summary, this paper makes the following contributions: 1) We propose a universal CDR model, \RC, which introduces motifs to capture the general topology knowledge across domains that can be applied to both intra-domain and inter-domain CDR tasks. 
2) We train \sRC under the ``Pre-training \& Prompt Tuning'' paradigm to seamlessly transfer the general topology knowledge, which can address the mismatch issue between the training objectives of SSL and recommendation tasks.
3)  Experimental results show the superiority of \sRC against the state-of-the-art methods on both intra-domain and inter-domain CDR tasks.

%% file: related.tex
\section{Related Work}

\vpara{Cross-domain Recommendation (CDR).} CDR can be categorized into intra-domain and inter-domain CDR. Intra-domain CDR addresses the data sparsity issue by transferring rich information from other domains to improve the intra-domain recommendation performance for users with few interactions. Existing works mainly use clustering~\cite{conf/cikm/MorenoSRS12,conf/ecir/KanagawaKSTS19,conf/ijcai/YuanYB19,conf/www/LianZXS17}, label relevance~\cite{conf/www/LiuZH022}, active learning~\cite{conf/aaai/ZhaoPXZLY13}, matrix decomposition~\cite{conf/sigir/YangYYLC17,conf/cikm/RafailidisC17}, dual transfer networks ~\cite{conf/cikm/HuZY18,conf/ijcai/ZhuWCLZ20,conf/ijcai/0008T0ZNY21,conf/www/LiuZZLSX0X20,guo2022time,journals/tkde/GuoZCWY23} to achieve knowledge transfer. Inter-domain CDR involves recommending items from a different domain to cold-start users. This task is more challenging than intra-domain CDR, as the cold-start users have no interactions in the new recommendation scenario. To solve this problem, researchers design mapping functions~\cite{conf/wsdm/ZhuTLZXZLH22,conf/cikm/KangHLY19,conf/sigir/ZhuGZXXZL021} to transfer the information of the cold-start users from the source to the target domain. Recently, several universal CDR models that unify both intra-domain and inter-domain CDR are proposed. These models first capture the general knowledge across domains through domain-shared embeddings, and then transfer this knowledge through MTL or PF. However, they ignore the fact that the general structural topology across domains is also a type of general knowledge.  %To solve this problem, we propose using motif to capture general topology knowledge.

\vpara{Prompt Learning for Recommendation.} Prompt learning solves the mismatch between the training objectives of the pre-training and downstream tasks by unifying these tasks as a common template, so as to active the model's memory for related tasks~\cite{brown2020language, journals/corr/abs-2203-02155}. The design of prompt templates can be categorized as discrete natural language prompts~\cite{conf/emnlp/ShinRLWS20,conf/acl/GaoFC20}, continuous prompts~\cite{journals/corr/abs-2103-10385}, and bias adjustment strategies~\cite{conf/acl/LoganBWP0022}. Recently, researchers  apply prompt learning  to solve various types of recommendation tasks, including news~\cite{conf/sigir/ZhangW23}, explainable~\cite{journals/corr/abs-2202-07371}, fairness~\cite{conf/sigir/WuXZZ0ZL022} and personalized~\cite{conf/recsys/Geng0FGZ22,conf/ecir/SileoVR22,conf/kdd/SunZHWW22,conf/www/LiuY0023} recommendation. However, prompt learning has not yet been applied to universal CDR. To fill this gap, we propose prompt learning to transfer general knowledge across domains.

%% file: approach.tex
\section{Problem Statement}

Let $\mathcal{D}=\{ \mathcal{D}_1, \mathcal{D}_2, \cdots \}$  be the multiple domains, where each domain $\mathcal{D}_i = \{V_i, E_i  \}$ consists of a node set $V_i = \{U_i, I_i \}$ and an edge set $E_i$. $U_i$ and $I_i$ denote the user and item set, respectively. CDR aims to improve the recommendation performance in all domains based on $\mathcal{D}$.
Notably, following UniCDR~\cite{conf/wsdm/CaoL0GL023}, \sRC supports both intra-domain and inter-domain CDR tasks that contain Dual-User-Intra, Dual-User-Inter, Multi-Item-Intra and Multi-User-Intra tasks, where Dual/Multi means whether the number of domains is two or more; User/Item means whether the users or items serve as the
overlapped role to bridge different domains; Intra/Inter means the corresponding CDR task. For convenience, we take the Dual-User-Intra task that contains two domains $\mathcal{D}=\{ \mathcal{D}_1, \mathcal{D}_2\}$   as an example to illustrate the workflow of \RC, and the other three tasks can be explained in a similar way.

\begin{figure*}[t]
	\centering
	\mbox{ 
		\subfigure[\scriptsize The overall workflow of \RC. ]{\label{subfig:overview1}
			\includegraphics[width=0.26\textwidth]{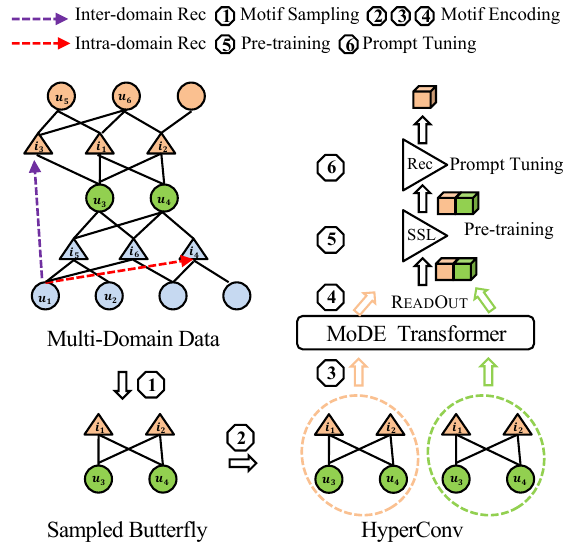}
		}

		\subfigure[\scriptsize The unified SSL and recommendation tasks, where the left is SSL within domain, the bottom is SSL across domains and the right is recommendation task.]{\label{subfig:overview2}
			\includegraphics[width=0.605\textwidth]{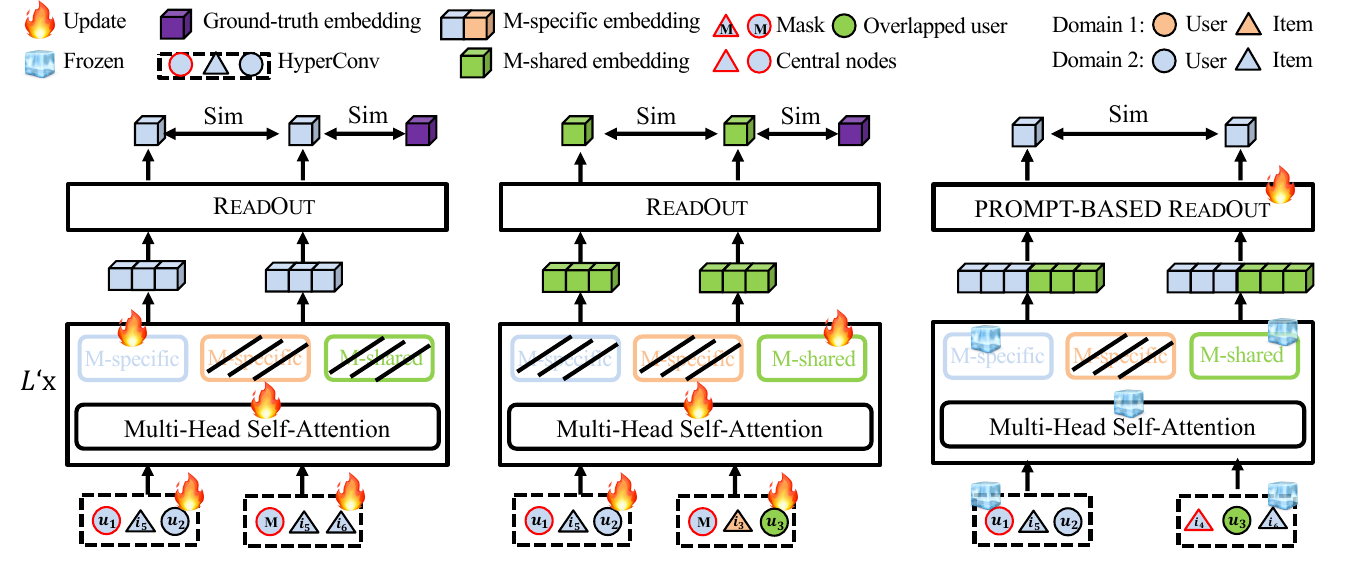}
		}

	}

	\caption{\label{fig:overview} \small{The whole framework of \RC.  Figure~\ref{subfig:overview2} shows an intra-domain CDR example, i.e., recommending $i_4$ to $u_1$ within the domain. For inter-domain CDR, i.e., recommending $i_3$ to $u_1$, we only use  M-shared embeddings for $u_1$, but use both M-shared and M-specific embeddings for $i_3$ to perform the recommendation task.}}
\end{figure*}

\section{\RC: Motif-based Prompt Learning }

Fig.~\ref{subfig:overview1} illustrates the workflow of \RC, which includes motif sampling and encoding, pre-training and prompt tuning processes. In this section, we first introduce
motif to capture the general structural topology across domains, as motif~\cite{milo2002network} involves the structural correlation of nodes.
Then we present a \textit{Motif-based Encoder} which consists of a hypergraph-based motif encoding module, a Mixture-of-Domain-Experts (MoDE) Transformer and a $\textsc{ReadOut}$ function
to obtain M-specific and M-shared embeddings. The hypergraph module models complex correlations of nodes, the MoDE Transformer routes the M-shared or M-specific experts for further embedding refinement, and the $\textsc{ReadOut}$ function introduces the central node signals to obtain the final node embeddings containing general structural topology information.
Next, we present the reformulated SSL and recommendation tasks into the same template that aims to narrow down the training objective gaps between them. Finally, we illustrate the prompt tuning technique that transfers the general knowledge into each domain to perform the downstream recommendation task.

\begin{figure}[t]
	\centering
	\mbox{ 
		\subfigure[\scriptsize The proposed three types of motifs.]{\label{subfig:motif}
			\includegraphics[width=0.282\textwidth]{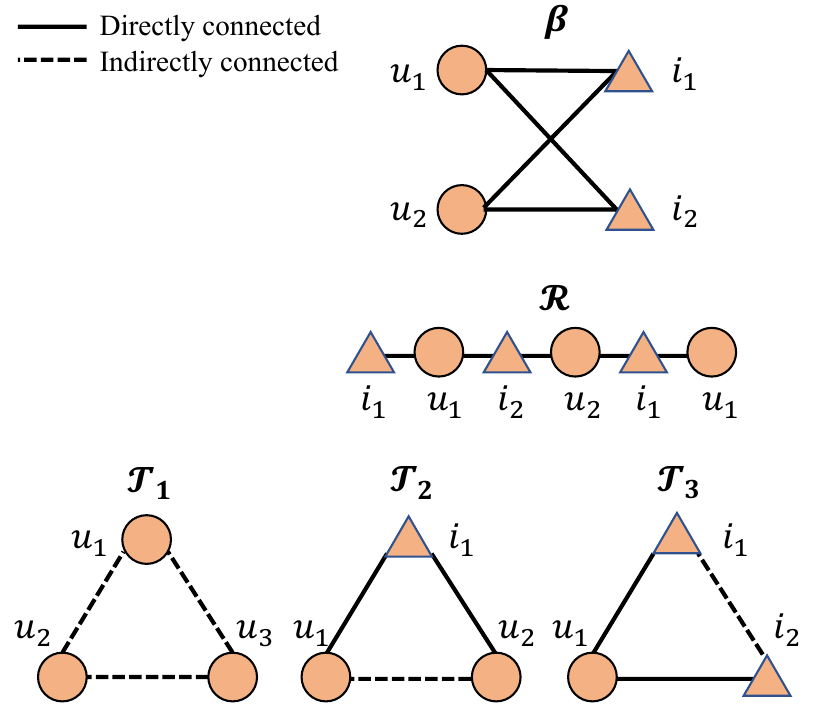}
		}
	
	\hspace{0.2cm}

		\subfigure[\scriptsize MoDE Transformer.]{\label{subfig:MoDE}
			\includegraphics[width=0.140\textwidth]{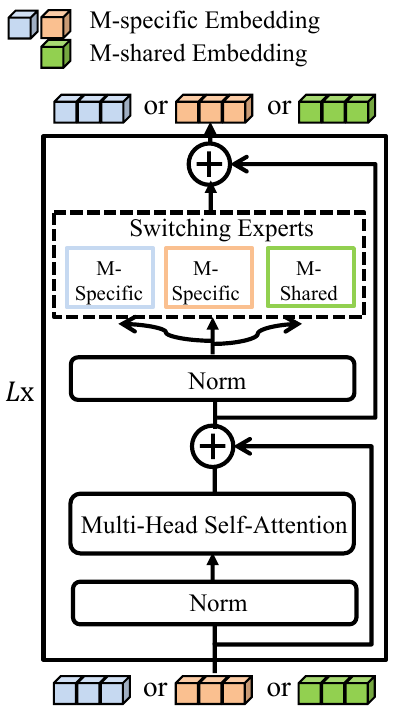}
		}

	}
	
	\caption{\label{fig:key_components} Key components in \RC.  }
\end{figure}

\hide{
\begin{figure}[t]
	\centering
	\includegraphics[width= .4 \textwidth]{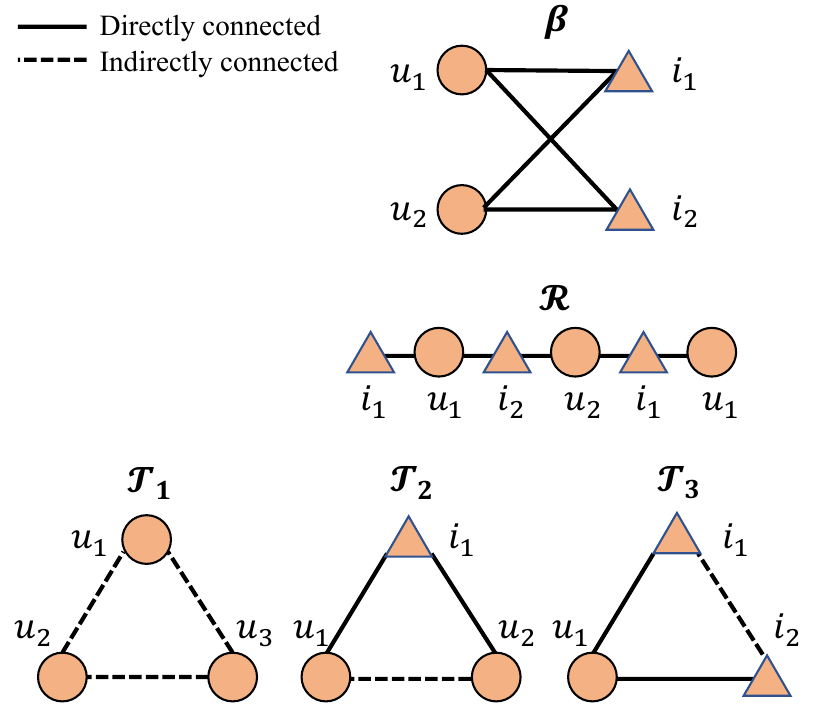}
	\caption{\label{fig:motif} The proposed three types of motifs.
	}
\end{figure}

\begin{figure}[t]
	\centering
	\includegraphics[width= .20 \textwidth]{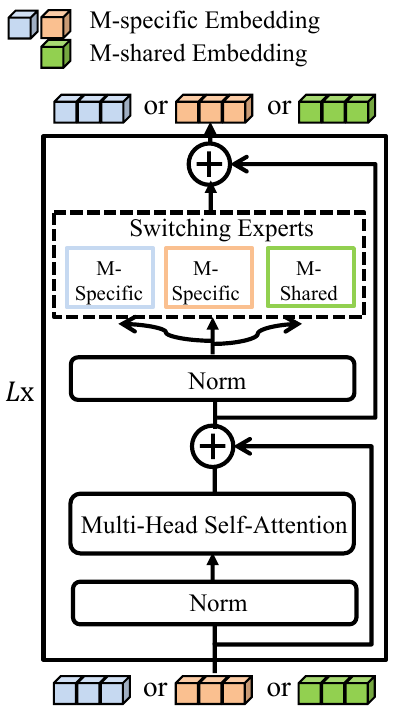}
	\caption{\label{fig:MoDE} MoDE Transformer with shared parameters.
	}
\end{figure}
}

\subsection{Motif Sampling}
{\label{subsec:motif_sampling}}
As motifs have been proven useful in capturing the general structural topology of graphs, we leverage them to model the common correlations of nodes across different domains, where users' interactions within a domain are represented as a user-item graph. By integrating domain knowledge into common motifs and sharing them across or within domains, we can enhance the recommendation performance in both inter-domain and intra-domain CDR tasks.
We present three types of motifs (Cf. Fig.~\ref{subfig:motif}), i.e., random walk~\cite{conf/sigir/CaoCLW22}, butterfly~\cite{conf/kdd/Sanei-MehriST18} and triangle~\cite{conf/www/YuYLWH021}, to respectively capture the long-term, fully connected short-term, and partially connected short-term dependencies of nodes in both inter-domain and intra-domain CDR tasks.
Notably,  our model can be seamlessly extended to
handle other motifs. 
We present the sampling process in domain $\mathcal{D}_1$ as an example, and the sampling process in domain $\mathcal{D}_2$ can be explained similarly.
 
 \subsubsection{Random Walk  Sampling}
Random walk, which has been extensively studied in the field of social network mining~\cite{conf/kdd/PerozziAS14}, has been successfully utilized to capture the long-term dependencies of nodes.
Formally, in domain $\mathcal{D}_i = \{ V_i, E_i\}$, a random walk $\mathcal{R} = \{n_1, n_2, \cdots, n_{|\mathcal{R}|}  \}$   is generated by the random walk algorithm~\cite{conf/kdd/PerozziAS14} upon the user-item graph, where $|\mathcal{R}|$ is the sequence length. The time complexity of random walk sampling is $\mathcal{O}((|V_i|* | \mathcal{R}| )$.

\subsubsection{Butterfly Sampling} 
Butterfly, which is a complete bipartite subgraph consisting of two nodes of one type and two nodes of another type~\cite{conf/kdd/Sanei-MehriST18}, is capable of capturing the  fully connected short-term dependencies of nodes with strong connections.
Formally, in domain $\mathcal{D}_i = \{U_i, I_i, E_i \}$, a butterfly $\mathcal{B}$ is generated by four nodes  $u_1, u_2, i_1, i_2$, as shown in Fig.~\ref{subfig:motif}. Here  $\{u_1, u_2\} \in U_i$, $\{i_1, i_2\} \in I_i$, $u_1, u_2$ are all connected to $i_1, i_2$. Inspired by BFC-VP~\cite{journals/pvldb/WangLQZZ19}, which counts the number of butterflies based on the node priority, we propose a \underline{P}riority-based \underline{B}utterfly \underline{S}ampling method PBS.  We first introduce the concept of priority and then provide a detailed explanation of the PBS method.

\vpara{Definition1: Priority.} In domain $\mathcal{D}_i = \{U_i, I_i, E_i \}$, the priority $p(n)$  of each node $n \in \{U_i, I_i \} $ is an integer where $p(n) \in [1, |U_i| + | I_i  |]$. For any two nodes $n_1, n_2 \in \{U_i, I_i\}$, the condition $p(n_1) > p(n_2)$ is satisfied if either 1) $deg(n_1) > deg(n_2)$  or 2) $deg(n_1) = deg(n_2)$ and $n_1.id > n_2.id$. Here, $n_1.id$ and $n_2.id$ represent the indices of $n_1$ and $n_2$, $deg(n_1)$ and $deg(n_2)$ represent the  degrees of these nodes.
Since domain $\mathcal{D}_i$ is an undirected graph, the priority of each node $n$ can be calculated simply by determining the number of its neighbor sets, i.e., $|\mathcal{N}(n)|$, where $|\mathcal{N}(n)|$ denotes the first-order neighbor set of node $n$.

Algorithm~\ref{algo:bts} shows the details of our PBS method. Given $\mathcal{D}_i = \{U_i, I_i, E_i \}$, we first calculate the priority $p(n)$ of each central node $n$, and then sort them in descending order. After that, for each central node $n$ in the sorted list, we find its second-order set $\{\mathcal{N}(n') | n' \in \mathcal{N}(n) \}$ and calculate the common neighbor set $\mathcal{N}(n) \cap \mathcal{N}(n'')$ between the central node $n$ and its second-order neighbor $n''$.
We then select two nodes from this common neighbor  set (e.g., $n_1, n_2 \in \mathcal{N}(n) \cap \mathcal{N}(n'') $), together with node $n, n''$, form a butterfly. Notably, following~\cite{journals/pvldb/WangLQZZ19},  to avoid duplication, we process the nodes that meet $p(n'')$<$ p(n')$<$ p(n)$.
Besides, the process of selecting two nodes from the common neighbor set can be viewed as a combination problem~\cite{journals/JJL02}, where we should enumerate two nodes from the set  $\mathcal{N}(n) \cap \mathcal{N}(n'')$.
The total time complexity of PBS is $\mathcal{O}(  \sum_{	(n, n')\in E_i}  (min\{deg(n), deg(n')\} )^2)$, as the process of finding the common neighbor set  is  $\mathcal{O} (\sum_{(n, n')\in E_i}   (  min\{deg(n), deg(n'')\} )^2 )$ (Line 1-10), and the time complexity of enumerating the  subsets  for all central nodes is also $\mathcal{O} (\sum_{(n,n')\in E_i} (min\{deg(n), deg(n'')\} )^2) $ (Line 11-14).

			\normalem
\begin{algorithm}[t]
	{\small \caption{ PBS: The Sampling Process of Butterfly.  \label{algo:bts}}
		\KwIn{$\mathcal{D} _i = \{U_i, I_i, E_i  \}$, The empty butterfly dictionary $\mathcal{B}_i$.}
		\KwOut{Sampled butterfly dictionary $\mathcal{B}_{i}$.}	
		\ForEach{$n \in  \{U_i, I_i\}$}{Compute $p(n)$\; Sorted $\mathcal{N}(n)$ in descending order according to  $p(n)$;}	
		
		\ForEach{$n \in  \{U_i, I_i\}$ }{ cur = [] \;
			\ForEach{ $n' \in \mathcal{N}(n): p(n') < p(n)$}{
				\ForEach {$n'' \in \mathcal{N}(n'): p(n'') < p(n)$}{ \If{$ \{n, n''  \} \notin \mathcal{B}_i.{\rm keys()}$ }{
													$\mathcal{B}_i[\{n, n''\}] = []$ \;
													cur.append([$n, n''$]) \;
				 			}
				}
			}
		
			\ForEach{ $\{n, n''\} \  in  \ \  {\rm cur}$ }
								{
									Select all possible subsets $s_i$ = [$n_{i_1}, n_{i_2}$] in $\mathcal{N}(n) \cap \mathcal{N}(n'')$ \;
									\ForEach{$s_i = [n_{i_1}, n_{i_2}]  \in \mathcal{N}(n) \cap \mathcal{N}(n'')$}{$\mathcal{B}_i[\{n, n''\}].{\rm append}([n_{i_1}, n_{i_2}])$ \;}
		
								}

		}
	\Return{$\mathcal{B}_i$}
	
	}
\end{algorithm}
\ULforem

\subsubsection{Triangle Sampling}
 Triangle is successfully deployed to capture the partially connected short-term dependencies of nodes with weak connections~\cite{conf/www/YuYLWH021,journals/science.aad9029}. 
Formally,  in domain $\mathcal{D}_i = \{U_i, I_i, E_i \}$, a triangle is defined as a set of three nodes  $\mathcal{T} = \{n_1, n_2, n_3\}$, where some nodes are  indirectly connected (e.g., users are friends with each other). 
As shown in Fig.~\ref{subfig:motif}, we propose three types of triangles. $\mathcal{T}_1 \subset \mathcal{T}$ depicts three users are friends with each other, where any two users contain at least $a_1$ common items. $\mathcal{T}_2 \subset \mathcal{T}$ depicts two similar users purchasing one item, where the users contain at least $a_2$ common items. 
$\mathcal{T}_3 \subset \mathcal{T}$ depicts a user and his interacted items, where the item similarity should be greater than $a_3$. 

In practice, we set the values of $a_1$ and $a_2$ as the average median number of items that the users in each domain interact with. As $a_3$ represents the similarity between items, we use ${\rm EASE}^R$~\cite{conf/www/Steck19} to calculate the item-item matrix $\mathbf{B}$, and set $a_3$ as 0 to represent the lowest similarity score. Following ~\cite{conf/www/Steck19}, we first obtain the adjacent matrix $\mathbf{A} \in \{0, 1\}^{| U_i | \times |I_i| }$ based on $\mathcal{D}_i = \{U_i, I_i, E_i \}$, then calculate $\mathbf{B}$ through $\mathbf{B}=\mathbf{I}-\mathbf{P} \cdot \operatorname{\textsc{DiagMat}}(\mathbf{1} \oslash \operatorname{\textsc{Diag}}(\mathbf{P}))$, where $\mathbf{P}=\left(\mathbf{A}^{\top} \mathbf{A}+\lambda_{F} \mathbf{I}\right)^{-1}$,  $\lambda_{F}$ is a hyperparameter of Frobenius norm regularizer, $\textsc{Diag}(\cdot)$ means the matrix diagonal elements, $\operatorname{\textsc{DiagMat}}$ is a diagonal matrix, $\mathbf{I}$ is identity matrix, $\mathbf{1}$ is a vector of ones and  $\oslash$ means element-wise division.
The time complexity of  sampling $\mathcal{T}_1, \mathcal{T}_2$  is both $\mathcal{O}(\sum_{n \in U_i}  {\rm deg}(n)  ) $, while  the time complexity of sampling $\mathcal{T}_3 
$ is $\mathcal{O}(|I_i|^2) + \mathcal{O}(\sum_{n \in U_i}  {\rm deg}(n)  ) $.

For the sake of convenience, in \textbf{~\secref{sec:motif_based_encoder}-~\secref{sec:downstream}}, we take a triangle as an example to illustrate the encoding process,
and other motifs can be explained in a similar way.

\subsection{Motif-based Encoder}
This section introduces the motif-based encoder, which consists of a hypergraph-based motif encoding module, a MoDE Transformer and a $\textsc{ReadOut}$ function.

{\label{sec:motif_based_encoder}}
\subsubsection{Hypergraph Convolution}
We apply hypergraph convolution upon the sampled motifs to refine node embeddings, as hypergraph~\cite{bretto2013hypergraph} brings a natural way to model complex correlations of nodes.
Following~\cite{conf/www/YuYLWH021}, we define our hypergraph convolution as: $\mathbf{X}^{(l+1)}=\mathbf{D}^{-1} \mathbf{H} \mathbf{WB^{-1}} \mathbf{H}^{\mathrm{T}} \mathbf{X}^{(l)}$,  where $\mathbf{X}^{(l)}$ represents the embedding matrix of the whole node set  at the $l$-th convolution process, $\mathbf{D}$ and $\mathbf{B}$ are the  diagonal matrices, $\mathbf{H}$ is the incidence matrix, where $H_{i \epsilon}$ = 1 if the hyperedge $\epsilon$ contains a node $v_i \in \{U_i, I_i \}$, and 0 otherwise.

After $L$ convolution operations, we can obtain the refined node embedding matrix $\mathbf{X}^L=\frac{1}{L+1}\sum_{l=0}^{L}\mathbf{X}^{(l)}$.
To capture the general topology of each triangle across domains and the domain-specific topology within domains, we conduct the lookup operation twice in $\mathbf{X}^L$. This results in the acquisition of the M-shared embedding $\mathbf{T}^{0}_{S}$ and the M-specific embedding $\mathcal{T}^{0}_{\mathcal{D}_1}$, which are then fed into the MoDE Transformer.

\subsubsection{MoDE Transformer} 
{\label{sec:MoDE}}
To effectively capture the structural topology within or across domains, we propose the \underline{M}ixture-\underline{o}f-\underline{D}omain-\underline{E}xperts (MoDE) Transformer (c.f., Fig.~\ref{subfig:MoDE}), which is inspired by the MoME Transformer~\cite{conf/nips/BaoW0LMASPW22}, due to its ability to separately encode text and image domains, thereby facilitating effective interaction between multi-modalities. The MoDE Transformer encodes the topology information within or across domains by replacing the feed-forward networks in the standard Transformer~\cite{conf/nips/VaswaniSPUJGKP17} with M-shared or M-specific experts.
Formally, after performing hypergraph convolution, we feed the M-shared and M-specific embeddings $\mathbf{T}^{0}_{S} $, $\mathbf{T}^{0}_{\mathcal{D}_1}$ into the MoDE Transformer. Take $\mathbf{T}^{0}_{S} $ as an example, the encoding process is: $\mathbf{T}^{l'}_{S} = {\rm MSA \ }( {\rm LN} \ (\mathbf{T}^{l-1}_{S})  )  +  \mathbf{T}^{l-1}_{S}$, and $	\mathbf{T}^{l}_{S} = {\rm MFFN \ }(  {\rm LN} \ (\mathbf{T}^{l'}_{S}) )   +   	\mathbf{T}^{l'}_{S}$, where $l$ is the $l$-th ($1 \leq l \leq L'$) Transformer layer, LN represents layer normalization, MSA denotes multi-head self-attention, and MFFN refers to routing experts.
At the final Transformer layer, we obtain the encoded triangle embedding, denoted as $\mathbf{T}^{L'}_{S}$. 
\subsubsection{\textsc{ReadOut} Function} Since the goal of recommendation is to obtain the central user/item embeddings,  to achieve this, we employ a $\textsc{ReadOut}$ function to aggregate the triangle embedding, and then 
incorporate the central node signal. Assuming the central node in $\mathbf{T}^{L'}_{S} $ is $n$, the process is defined as follows:

\beqn{
	\label{eq:aggregate}
	\mathbf{z}^{S} &=& \textsc{ReadOut}(\{\mathbf{t}^{L'}_{S}| \mathbf{t}^{L'}_{S} \in \mathbf{T}^{L'}_{S}  \}),  \\
	\label{eq:aggregate2}
	\mathbf{z}_n^{S} &=& \textsc{Concat}(  \mathbf{z}^S, \mathbf{z}_n'   ), 
}

\noindent where $	\mathbf{z}_n^{S}$ is the M-shared embeddings for node $n$, $\mathbf{z}^{S}$ denotes the shared aggregated motif embedding,  $\mathbf{z}_n' \in \mathcal{X}^L$ refers to the central node signal,  $ \mathbf{t}^{L'}_{S}$ is the refined node embedding in $\mathbf{T}^{L'}_{S}$, $\textsc{Concat}$ is the concatenate operation, and $\textsc{ReadOut}$ is instantiated as the mean operation. 
Similarly, we can obtain M-specific embeddings $\mathbf{z}_n^{\mathcal{D}_1}$ for node $n$.
	Next, we will illustrate how to train \sRC with both M-shared and M-specific embeddings under the ``Pre-training \& Prompt Tuning'' paradigm.

\subsection{Pre-training Stage}
{\label{sec:pretrain}}
To reduce the training objective gaps between the SSL and recommendation tasks, we unify them as a common template, namely the motif-based similarity learning (MSL) task.
We adopt contrastive learning (CL) and embedding reconstruction (ER) as the SSL tasks, but our model can also handle other SSL tasks.
Specifically, the CL task is reformulated as maximizing the similarity between motifs centered on the same node while minimizing the similarity between motifs centered on different nodes. 
The ER task is reformulated as maximizing the similarity between the ground-truth and predicted node embeddings.
To enhance interaction among different domains, both the CL and ER tasks are trained within and across different domains.
In the following part, we illustrate the unified common template and present the reformulated CL and ER tasks.

\vpara{Unified Task Template.}
We redefine the CL and ER tasks as the MSL task.
Formally, let $\mathbf{z}_{n}^{S}$ and $\mathbf{z}_{n}^{\mathcal{D}_1}$ represent the M-shared and M-specific  embeddings of node $n$, and let $\text{sim}(\cdot)$ denote the cosine similarity function. As shown in Fig.~\ref{subfig:overview2}, the CL and ER tasks can be mapped to the computation of motif-based node similarity, which is defined as follows.

\vpara{Contrastive Learning.} 
We maximize the similarity between motifs centered on the same node while minimizing the similarity between motifs centered by different nodes. 
To strengthen knowledge transfer among different domains, we not only sample motifs centered on nodes within each domain, but also sample motifs centered on overlapped nodes across domains.
Specifically,
the positive pair in the CL task consists of two parts, one for overlapped nodes, i.e.,  $\{ (\mathbf{z'}^{\mathcal{S}}_{n_s}, \mathbf{z''}^{\mathcal{S}}_{n_{s}} )| n_{s} \in   \{V_1 \cap V_2\} \} $, where $n_s$ is an overlapped node in domain $\mathcal{D}_1$ and $\mathcal{D}_2$; and one for non-overlapped nodes, denoted as 
$\{ (\mathbf{z'}^{\mathcal{D}_1}_{n_{p}}, \mathbf{z''}^{\mathcal{D}_1}_{n_p} )| n_{p} \in V_1\}$, where $n_p$ is sampled within $\mathcal{D}_1$. 
We treat  motifs centered on different nodes as negative pairs, denoted as $\{ (\mathbf{z'}^{\mathcal{D}_1}_{n_{p}}, \mathbf{z''}^{\mathcal{D}_1}_{n_q} )| p \neq q,   \{n_{p}, n_{q}\} \in   V_1  \}$.
Following SimCLR~\cite{conf/icml/ChenK0H20}, we use the InfoNCE loss~\cite{journals/jmlr/GutmannH10} to optimize the model parameters. The loss function in domain $\mathcal{D}_1$ is: 
\beqn{
	\label{eq:contrastive-learning-loss}
	\nonumber 
	&\mathcal{L}_{c_1} = - \sum_{n_{s} \in \{ V_1 \cap V_2 \} } \log \frac{\exp(\text{sim}(\mathbf{z'}^{S}_{n_{s}},\mathbf{z''}^{S}_{n_{s}} ) / \tau  )} {\sum_{n_q \in \{V_1 \}  \backslash n_{s}  }\exp(\text{sim}(\mathbf{z'}^{S}_{n_{s}} , \mathbf{z'}^{S}_{n_q} ) / \tau  )}  \\
	& - \sum_{n_{p} \in  V_1 \backslash \{ V_1 \cap V_2 \} } \log \frac{\exp(\text{sim}(\mathbf{z'}^{\mathcal{D}_1}_{n_{p}},\mathbf{z''}^{\mathcal{D}_1}_{n_{p}} ) / \tau  )} {\sum_{n_q \in \{V_1\} \backslash n_{p}  } \exp(\text{sim}(\mathbf{z'}^{\mathcal{D}_1}_{n_{p}}, \mathbf{z''}^{\mathcal{D}_1}_{n_q} ) / \tau  ) },   \\  \nonumber
}

\hide{
\beqn{
	\label{eq:contrastive-learning-loss}
	\nonumber 
	&\mathcal{L}_{c_1} = - \sum_{n_{p} \in  V_i  \backslash \{ V_i \cap V_j \} } \log \frac{\exp(\text{sim}(\mathbf{z}'_{n_{p}},\mathbf{z}''_{n_{p}} ) / \tau  )} {\sum_{n_q \in \{V_i \cup V_j\} \backslash n_{p}  } \exp(\text{sim}(\mathbf{z}'_{n_{p}}, \mathbf{z}''_{n_q} ) / \tau  ) } - \\ 
    & \sum_{n_{s} \in \{ V_i \cap V_j \} } \log \frac{\exp(\text{sim}(\mathbf{z}'_{n_{s}},\mathbf{z}''_{n_{s}} ) / \tau  )} {\sum_{n_q \in \{V_i \cup V_j\}  \backslash n_{s}  }\exp(\text{sim}(\mathbf{z}'_{n_{s}} , \mathbf{z}''_{n_q} ) / \tau  )},  \\  \nonumber
}
}

\noindent where  $\tau$ is the temperature hyperparameter. Similarly, we can define $\mathcal{L}_{c_2}$ in domain $\mathcal{D}_2$, and the total CL loss is $\mathcal{L}_c = \mathcal{L}_{c_1} + \mathcal{L}_{c_2}$.

\vpara{Embedding Reconstruction.} The objective of the ER task is to maximize the similarity between positive pairs while minimizing the similarity between negative pairs. Positive pairs consist of the predicted and ground-truth embeddings of the same node, whereas negative pairs consist of the predicted and ground-truth embeddings of different nodes.
Formally, we first sample a set of triangles $\mathcal{T}$ centered on $n$ with abundant interactions from $\{V_1  \cup V_2\}$, and use any CDR method (e.g., UniCDR~\cite{conf/wsdm/CaoL0GL023}) to obtain its ground-truth embedding $\mathbf{z}_{n}$, as Hao et al.~\cite{conf/wsdm/HaoZYL021,journals/tois/HaoYZLC23} have demonstrated the base recommendation model can generate high-quality node embeddings when the nodes have abundant interactions. We then mask some nodes, resulting in a masked triangle set $\mathcal{\hat{T}}$. Next, we randomly select one masked triangle and feed it into the Motif-based Encoder to obtain the reconstructed embedding. If $n$ is an overlapped node,  we feed the corresponding M-shared embedding into the Motif-based Encoder, resulting in a reconstructed embedding $\mathbf{\hat{z}}^{S}_{n}$. If $n$ is a non-overlapped node, we instead use M-specific embedding, resulting in a reconstructed embedding $\mathbf{\hat{z}}^{\mathcal{D}_1}_{n}$.
Similar to the CL task, we also adopt the InfoNCE loss to optimize the model parameters. The loss function is defined as follows:
\beqn{
	\label{eq:embedding-reconstruction-loss}
	\nonumber 
	&\mathcal{L}_{e_1} = - \sum_{n_{s} \in \{ V_1 \cap V_2 \} } \log \frac{\exp(\text{sim}( \mathbf{z}_{n_s}, \mathbf{\hat{z}}^{S}_{n_s} ) / \tau  )} {\sum_{n_q \in \{V_1 \}  \backslash n_{s}  }\exp(\text{sim}(\mathbf{z}_{n_s} , \mathbf{\hat{z}}^{S}_{n_q} ) / \tau  )}  \\
	& - \sum_{n_{p} \in  V_1 \backslash \{ V_1 \cap V_2 \} } \log \frac{\exp(\text{sim}(\mathbf{z}_{n_{p}},\mathbf{\hat{z}}^{\mathcal{D}_1}_{n_{p}} ) / \tau  )} {\sum_{n_q \in \{V_1\} \backslash n_{p}  } \exp(\text{sim}(\mathbf{z}_{n_{p}}, \mathbf{\hat{z}}^{\mathcal{D}_1}_{n_q} ) / \tau  ) }.   \\  \nonumber
}

\hide{
\beqn{
	\label{eq:embedding-reconstruction-loss}
	\mathcal{L}_{e} = - \sum_{n \in  \{V_i  \cup V_j\} } \log  \frac{ \exp( \text{sim} (\mathbf{\hat{z}}_{n}, \mathbf{z}_{n} ) )}{ \sum_{q \in \{ V_i \cup V_j \} \backslash  n } \exp( \text{sim} (\mathbf{\hat{z}}_{q}, \mathbf{z}_{n} ) )  } ,
}
\noindent Notably, for each target node $n$, we sample the negative node $q$  from the same  or different domain with $n$. In Eq.~\eqref{eq:embedding-reconstruction-loss}, if $n$ and $q$ from the same domain, we assign them embeddings from the domain-specific embedding matrix, and choose the domain-intra experts in the MoDE Transformer. Otherwise, we assign them embeddings from the shared embedding matrix, and choose the domain-inter experts in the MoDE Transformer.}

\noindent   Similarly, we can define $\mathcal{L}_{e_2}$ in domain $\mathcal{D}_2$. The total ER loss is  $\mathcal{L}_e = \mathcal{L}_{e_1} + \mathcal{L}_{e_2}$.  
The pre-training loss $\mathcal{L}_{pre}$ $ = \lambda_1 \mathcal{L}_c + (1-\lambda_1) \mathcal{L}_e$,  where  $\lambda_1$ controls the balance of the CL and ER tasks.

\subsection{Prompt Tuning Stage}
{\label{sec:downstream}}

The goal of prompt tuning is twofold: 1) unleash \RC's inherent capacity to generate high-quality user/item embeddings tailored for recommendation tasks within each domain, and 2) unify the  recommendation task into the common MSL task as outlined in Section~\ref{sec:pretrain}, so as to align the training objectives between the SSL and recommendation tasks.
%We add learnable parameters to the $\textsc{ReadOut}$ function in each domain to guide \sRC in generating M-shared and M-specific embeddings. These embeddings are then fed into a unified recommendation task that can be applied to both intra-domain and inter-domain CDR tasks. 
In the following part, we present the prompt design and illustrate the prompt tuning process in solving both the intra-domain and inter-domain CDR tasks.

\vpara{Prompt Design.}
In contrast to the prompt tuning methods in natural language processing that use handwritten prompt words to stimulate the model's understanding of different contexts~\cite{journals/corr/abs-2203-02155}, our focus is on capturing the topology in the user-item graph. 
As the $\textsc{ReadOut}$ function plays a key role in refining motif-induced embeddings for recommendation, we add learnable parameters to the $\textsc{ReadOut}$ function in Eq.~\eqref{eq:aggregate} and  $\mathbf{p}$ in Eq.~\eqref{eq:aggregate2} in each domain to guide \sRC in generating distinctive M-shared and M-specific embeddings for different downstream recommendation tasks.
%To this end, in each domain, we propose three types of prompt-based $\textsc{ReadOut}$ operations, where  we add  learnable prompt parameters $\mathbf{p}_n$  in Eq.~\eqref{eq:aggregate} and  $\mathbf{p}$ in Eq.~\eqref{eq:aggregate2} to play an auxiliary role in generating M-shared and M-specific embeddings.
Here we take generating the M-shared embedding $\mathbf{z}_{n}^{S}$ as an example, and the process of generating M-specific embedding $\mathbf{z}_{n}^{\mathcal{D}_1}$  in domain $\mathcal{D}_1$ can be explained similarly. The proposed $\textsc{ReadOut}$-based prompts are shown as follows:

\begin{itemize}[leftmargin=*]
	\item Element-wise Multiplication: \\
	\beqn{
		\label{eq:element-wise}
		\mathbf{z}^S= \textsc{ReadOut}(\{ \mathbf{p}_n  \odot     \mathbf{t}_S^{L'}| \mathbf{t}_S^{L'} \in \mathbf{T}_S^{L'}  \}),
	}	

	\noindent where $\mathbf{z}^S$ refers to the shared aggregated motif embedding,  $ \odot$ denotes the element-wise multiplication. This operation is  a feature weighted summation of the nodes in the motif.

	\item Matrix Multiplication: \\
		\beqn{
		\label{eq:matrix-multiplication}
		\mathbf{z}^S= \textsc{ReadOut}(\{ \mathbf{p}_n  \mathbf{t}_S^{L'}| \mathbf{t}_S^L \in \mathbf{T}_S^{L'}  \}),
	}
	
	\noindent  This operation is a linear projection upon the nodes in the motif.
	
	\item Attention:
			\beqn{
		\label{eq:attention}
		\mathbf{z}^S = \textsc{ReadOut}(\{ p_{n_j}  \mathbf{t}_{S_j}^{L'}| \mathbf{t}_{S_j}^{L'} \in \mathbf{T}_{S}^{L'}  \}),
	}
	
	\noindent where $p_{n_j} = \frac{\exp(\mathbf{W} \mathbf{t}_{S_j}^{L'})} {  \sum_{h=1}^{3}  \exp(\mathbf{W} \mathbf{t}_h^{L'})}   $ is the attention weight for the node embedding $\mathbf{t}_{S_j}^{L'}$ in the motif, $\mathbf{W}$ is a weight matrix. This operation can distinguish the importance of nodes in the motif.
\end{itemize}

After we obtain  $\mathbf{z}^{S}$,  we further add  $\mathbf{p}$ in Eq.~\eqref{eq:aggregate2} to obtain the M-shared  embedding $\mathbf{z}^S_n$, i.e., $	\mathbf{z}^S_n = \mathbf{p} \cdot \  \textsc{Concat}(  \mathbf{z}^S, \mathbf{z}_n'   ) $.  Similarly, we can obtain the M-specific embedding $\mathbf{z}^{\mathcal{D}_1}_n$.

\vpara{Prompt Tuning.} We froze the parameters of Motif-based Encoder and only update the prompt parameters in the unified recommendation task (As shown in Fig.~\ref{subfig:overview2}). The loss function in domain $\mathcal{D}_1$ is defined as follows:

\beqn{
	\label{eq:prompt-tuning}
	 \mathcal{L}_{Rec_1} = - \sum_{(u, i) \in E_i } \log \frac{\exp(\text{sim}(\mathbf{\tilde{z}}_{u},\mathbf{\tilde{z}}_{i} ) / \tau  )} {\sum_{(u. i') \notin E_i  } \exp(\text{sim}(\mathbf{\tilde{z}}_{u}, \mathbf{\tilde{z}}'_{i} ) / \tau  ) }.  \\ \nonumber
}

\noindent where $\tilde{z}_u$ and $\tilde{z}_i$ are refined user/item embeddings, which are capable of  performing  both intra-domain and inter-domain CDR tasks.
For intra-domain CDR,  we set $\tilde{z}_u$ as $\textsc{Concat}(\mathbf{z}^S_u, \mathbf{z}^{\mathcal{D}_1}_u)$, while for inter-domain  CDR, we set $\tilde{z}_u$ as $\mathbf{z}^S_u$. 
We set $\tilde{z}_i$ as $\textsc{Concat}(\mathbf{z}^S_i,  \mathbf{z}^{\mathcal{D}_1}_i)$ in both these two tasks.
Similarly, we can obtain the recommendation loss $\mathcal{L}_{Rec_2}$ in domain $\mathcal{D}_2$, and the total recommendation loss is $\mathcal{L}_{Rec}$ = $\mathcal{L}_{Rec_1}$ + $\mathcal{L}_{Rec_2}$.

%% file: exp.tex
\section{Experiments}
We conduct comprehensive experiments on four CDR scenarios to answer the following research questions: \textbf{RQ1:} How does our proposed \sRC model perform when compared to the state-of-the-art CDR models? \textbf{RQ2:} Are SSL tasks (i.e., the CL and ER tasks) beneficial to the downstream \underline{rec}ommendation (\textbf{Rec}) task?  \textbf{RQ3:} What are the impacts of the three proposed  motifs on the downstream Rec task? \textbf{RQ4:} Are the encoding components (i.e., hypergraph and MoDE Transformer)  beneficial to the downstream Rec task? \textbf{RQ5:} Does the proposed ``\underline{P}re-training \& \underline{P}rompt \underline{T}uning'' (\textbf{PPT}) paradigm, which includes prompt tuning and the prompt-based $\textsc{ReadOut}$ function, benefit the downstream Rec task? \textbf{RQ6:} How do the hyperparameters of \sRC affect the downstream Rec task?

\subsection{Datasets}

Following UniCDR~\cite{conf/wsdm/CaoL0GL023}, we conduct four CDR tasks, including Dual-User-Intra (Scenario 1), Dual-User-Inter (Scenario 2), Multi-Item-Intra (Scenario 3) and Multi-User-Intra (Scenario 4). The statistics of the datasets utilized in these tasks are presented in Table~\ref{tb:datasets}.

\begin{table}[t]
	\newcolumntype{?}{!{\vrule width 0.2pt}}
	\caption{
		\label{tb:datasets} Statistics of four CDR scenarios.
		\normalsize
	}
	\centering  \small
	\renewcommand\arraystretch{1.0}
	\resizebox{.46 \textwidth}{!}{
	\begin{tabular}{c?c?c?c?c?c?c?c}
		\hline   {\centering Scenarios}  & {\centering Datasets}  & {\centering $|U|$} & $|V|$ &  {\centering Training}  &  {\centering Valid} &  {\centering Test}  & {\centering Median} \\
		\hline \multirow{5}{*}[0.8ex]{ Scenario 1 } &  Sport  &  9,928 &  30,796 & 92,612 & - & 8,326  & 6\\
		&    Cloth & 9,928 & 39,008 & 87,829 & - & 7,540  & 6\\
		\cline { 2 - 8 } &    Elec & 3,325 & 17,709 & 50,407 & - & 2,559 & 21 \\
		&  Phone  & 3,325 & 38,706 & 115,554 & - & 2,560  &12\\
		\hline \hline \multirow{5}{*}[0.5ex]{\centering Scenario 2}
		 &   Sport  & 27,328 & 12,655 & 163,291 & 3,589 & 3,546 & 5 \\
		&  Cloth  & 41,829 & 17,943 & 187,880 & 3,156 & 3,085 & 4 \\
		\cline { 2 - 8 } &   Game  & 25,025 & 12,319 & 155,036 & 1,381 & 1,304  & 5\\
		&   Video  & 19,457 & 8,751 & 156,091 & 1,435 & 1,458 & 6\\
		\hline \hline \multirow{5}{*}[-0.4ex]{\centering Scenario 3} 
		&   M1  & 7,109 & 2,198 & 48,302 & 3,526 & 3,558  & 6\\
		&   M2  & 2,697 & 1,357 & 19,615 & 1,362 & 1,310  & 6 \\
		&   M3  & 3,328 & 1,245 & 23,367 & 1,629 & 1,678 & 6 \\
		&   M4 & 5,482 & 2,917 & 41,226 & 2,720 & 2,727  & 6\\
		&   M5  & 6,466 & 9,762 & 77,173 & 3,090 & 3,154 & 10 \\
		\hline \hline \multirow{5}{*}[1.8ex]{ Scenario 4 }
		 &   D1  &  231,444& 2,096& 491,098&13,435 &13,437  & 1\\
		 &   D2  & 507,715 & 595 & 1,068,490 & 36,013 & 35,985 & 1\\
		 &   D3  & 773,188 & 1,312 & 3,785,720 & 92,659 & 92,672 & 3\\
		\hline
	\end{tabular}
}
\end{table}

\subsection{Experimental Setting}
\subsubsection{Evaluation Protocol}
 We adopt  HR@$K$ and NDCG@$K$ as the evaluation metrics, with $K$ set to 10 by default. Following previous study~\cite{conf/wsdm/CaoL0GL023}, for the first three scenarios, we conduct the leave-one-out strategy, i.e., for each positive item in the valid/test set, we randomly sample 999 negative items to alleviate the biased Rec phenomena~\cite{conf/kdd/KricheneR20}. We adopt the full-ranked strategy in the fourth CDR scenario to evaluate the model performance.

\subsubsection{Baselines}

We compare \sRC with universal CDR models including UniCDR~\cite{conf/wsdm/CaoL0GL023}, CAT-ART~\cite{conf/wsdm/LiXYHLSQN23}, $\textsc{GraphPrompt}$~\cite{conf/www/LiuY0023} and GPPT~\cite{conf/kdd/SunZHWW22}. We also select other  baseline models presented in UniCDR~\cite{conf/wsdm/CaoL0GL023}. For ease of comparison, we report the best performance of the baseline models and use the symbol ``SOTA'' to denote it.

\subsubsection{Implementation Details} For fair comparison, we follow the original settings for the baseline models. We set the latent embedding dimension as 128, the learning rate as 0.001, the batch size as 1024, the negative number as 4, the random walk length as \{3, 6, 9\}, the temperature $\tau$ as 0.5, the balance factor $\lambda_1$ as 0.5, the MoDE Transformer layer $L'$ as 2, the hypergraph layer $L$ as 4, $a_3$ as 0 in all CDR scenarios. Table~\ref{tb:datasets} shows the settings of $a_1$, $a_2$ in all CDR scenarios.
We use the butterfly as the motif and element-wise multiplication as  the $\textsc{ReadOut}$ function by default.
\subsection{Performance Comparisons}
\subsubsection{Overall Performance (RQ1)}

We report the overall Rec performance on four CDR  scenarios in Table~\ref{tb:dual_user_intra_inter} and Table~\ref{tb:multi_user_item_intra}. The results show that: 1) our proposed \sRC has the best or most competitive performance than the ``SOTA'' methods. 2) \sRC supports both  inter-domain and inter-domain CDR tasks, which allows for more efficient use of time and resources in the model design process.

\begin{table}[t]
	\newcolumntype{?}{!{\vrule width 0.5pt}}
	\scriptsize
	\centering
	\caption{Performance comparison (\%) on Dual-User-Intra/Dual-User-Inter CDR tasks. \small{$a$: UniCDR~\cite{conf/wsdm/CaoL0GL023}, $b$: DisenCDR~\cite{conf/sigir/CaoLCYL022}, $c$: CDRIB~\cite{conf/icde/CaoSCLW22}}}
	\setlength\tabcolsep{2.1pt}{
		\begin{tabular}{cc?cccc?cccc}
			\hline
			\multirow{2}{*}{\makecell*[c]{Models}} & \multirow{2}{*}{\makecell*[c]{Metrics}} & \multicolumn{4}{c?}{\makecell*[c]{Dual-User-Intra}} &  \multicolumn{4}{c}{Dual-User-Inter}  \\  
			\cline{3-10}  & & \makecell*[c]{Sport} &\makecell*[c]{Cloth} & Elec & Phone & Sport & Cloth & Game & Video \\ 
			\hline
			\multirow{2}{*}{\makecell*[c]{SOTA}}     & HR@10    & ${\makecell*[c]{18.37}}^a$    &${\makecell*[c]{17.85}}^a$    & ${\makecell*[c]{24.57}}^b$ &${\makecell*[c]{28.76}^b}$&  ${\makecell*[c]{12.04}}^c$ &${\makecell*[c]{12.48}}^a$ &${\makecell*[c]{\textbf{8.78}}}^a$& ${\makecell*[c]{13.17}}^c$ \\
			\multirow{2}{*}{}          & NDCG@10                        & ${\makecell*[c]{10.98}}^a$    & ${\makecell*[c]{\textbf{11.20}}}^a$ & ${\makecell*[c]{14.51}}^b$ & ${\makecell*[c]{16.13}^b}$&${\makecell*[c]{7.04}}^a$&${\makecell*[c]{7.52}}^a$ &${\makecell*[c]{\textbf{4.63}}}^a$&${\makecell*[c]{6.49}}^c$ \\
			\cline{1-10}
			%\specialrule{0em}{2pt}{2pt}
			\hline
			\multirow{2}{*}{\makecell*[c]{\RC}}     & HR@10   &   ${\makecell*[c]{\textbf{19.23}}}$    & ${\makecell*[c]{\textbf{18.18}}}$&${\makecell*[c]{\textbf{25.19}}}$&${\makecell*[c]{\textbf{28.82}}}$&   ${\makecell*[c]{\textbf{13.09}}}$ &  ${\makecell*[c]{\textbf{13.03}}}$ &  ${\makecell*[c]{7.98}}$ &  ${\makecell*[c]{\textbf{14.13}}}$  \\
			\multirow{2}{*}{}          & NDCG@10                        &      ${\makecell*[c]{\textbf{11.37}}}$     &${\makecell*[c]{10.19}}$           & ${\makecell*[c]{\textbf{15.11}}}$        &${\makecell*[c]{\textbf{18.02}}}$     &${\makecell*[c]{\textbf{8.09}}}$ &${\makecell*[c]{\textbf{9.01}}}$ &${\makecell*[c]{4.09}}$ &${\makecell*[c]{\textbf{6.88}}}$   \\
			\hline
		\end{tabular}
	}
	\hide{
		\begin{center}
			\hide{ \normalsize * indicates that the improvements are statistically significant for $p < 0.05$ judged with the runner-up result in each case by paired t-test.}
			$a$: UniCDR~\cite{conf/wsdm/CaoL0GL023}, $b$: DisenCDR~\cite{conf/sigir/CaoLCYL022}, $c$: CDRIB~\cite{conf/icde/CaoSCLW22}
		\end{center}
	}
	\label{tb:dual_user_intra_inter}
\end{table}

\begin{table}[t]
	\newcolumntype{?}{!{\vrule width 0.5pt}}
	\scriptsize
	\centering
	\caption{Performance comparison (\%) on Multi-Item-Intra/Multi-User-Intra CDR tasks. \small{$a$: UniCDR~\cite{conf/wsdm/CaoL0GL023}, $d$: ${\rm S}^3$-Rec~\cite{conf/cikm/ZhouWZZWZWW20}}}
	\setlength\tabcolsep{1.7pt}{
		\begin{tabular}{cc?ccccc?ccc}
			%\toprule
			\hline
			\multirow{2}{*}{Models} & \multirow{2}{*}{Metrics} & \multicolumn{5}{c?}{Multi-Item-Intra} &  \multicolumn{3}{c}{Multi-Item-Intra}  \\  
			\cline{3-10}  & & \makecell*[c]{M1} & M2 & M3 & M4 & M5 & D1 & D2 & D3 \\ 
			%\midrule
			\hline 
			\multirow{2}{*}{SOTA}     & HR@10  & ${\makecell*[c]{73.13}}^d$    & ${\makecell*[c]{60.86}}^d$   & ${\makecell*[c]{66.53}}^d$  & ${\makecell*[c]{48.46}}^d$&${\makecell*[c]{22.66}}^d$& ${\makecell*[c]{\textbf{32.60}}}^a$& ${\makecell*[c]{64.37}}^a$& ${\makecell*[c]{73.89}}^a$\\
			\multirow{2}{*}{}          & NDCG@10   & ${\makecell*[c]{59.57}}^a$  & ${\makecell*[c]{47.52}}^a$   &  ${\makecell*[c]{\textbf{53.24}}}^a$ &${\makecell*[c]{42.54}}^a$&${\makecell*[c]{18.63}}^d$& ${\makecell*[c]{\textbf{13.56}}}^a$& ${\makecell*[c]{50.48}}^a$& ${\makecell*[c]{59.15}}^a$ \\ 
			\cline{1-10} 
			%\specialrule{0em}{2pt}{2pt}
			% \midrule
			\hline
			\multirow{2}{*}{\RC}     & HR@10                          &${\makecell*[c]{\textbf{74.01}}}$ & ${\makecell*[c]{\textbf{61.29}}}$  & ${\makecell*[c]{\textbf{68.03}}}$ &${\makecell*[c]{\textbf{49.02}}}$ &${\makecell*[c]{\textbf{22.79}}}$ &${\makecell*[c]{31.88}}$ &${\makecell*[c]{\textbf{65.24}}}$ &${\makecell*[c]{\textbf{74.29}}}$ \\
			\multirow{2}{*}{}          & NDCG@10     & ${\makecell*[c]{\textbf{60.10}}}$   & ${\makecell*[c]{\textbf{48.01}}}$  &${\makecell*[c]{52.08}}$  &${\makecell*[c]{\textbf{43.82}}}$ &${\makecell*[c]{\textbf{20.06}}}$ &${\makecell*[c]{12.86}}$&${\makecell*[c]{\textbf{66.23}}}$&${\makecell*[c]{\textbf{60.02}}}$ \\
			%\bottomrule
			\hline
		\end{tabular}
	}
	
	\hide{
		\begin{center}
			$a$: UniCDR~\cite{conf/wsdm/CaoL0GL023}, $d$: ${\rm S}^3$-Rec~\cite{conf/cikm/ZhouWZZWZWW20}
		\end{center}
	}
	
	\label{tb:multi_user_item_intra}
\end{table}

\begin{table}[t]
	\newcolumntype{?}{!{\vrule width 0.5pt}}
	\scriptsize
	\centering
	\caption{Performance comparison (\%) of pre-training tasks.}
	\setlength\tabcolsep{5.2pt}{
		\begin{tabular}{lc?ccc}
			%\toprule
			\hline
			\multirow{2}{*}{\makecell*[c]{Scenarios}} & \multirow{2}{*}{Metrics} & \multicolumn{3}{c}{\makecell*[c]{Variant Models}}     \\  
			\cline{3-5}  & & \makecell*[c]{CL-/CL*/CL} & ER-/ER*/ER & A-/A*/A  \\
			%\midrule 
			\hline
			\multirow{2}{*}{\makecell*[c]{Sport}}     & HR@10                &\makecell*[c]{10.37/12.23/18.45} & \makecell*[c]{8.84/11.28/13.25} &    \makecell*[c]{11.03/18.89/\textbf{19.23}}                 \\
			\multirow{2}{*}{}   & NDCG@10  & 6.37/8.23/10.49      &\makecell*[c]{5.73/7.74/9.45}& \makecell*[c]{9.37/11.19/\textbf{11.37}}                    \\
			\cline{2-5}
			%\specialrule{0em}{2pt}{2pt}
			% \midrule
			\hline
			\multirow{2}{*}{Cloth}     & HR@10  & \makecell*[c]{15.88/16.74/17.99} &\makecell*[c]{6.72/8.48/10.42}&  \makecell*[c]{14.37/17.98/\textbf{18.18}}                         \\
			\multirow{2}{*}{}          & NDCG@10    &   \makecell*[c]{8.68/8.91/9.92}      &\makecell*[c]{5.64/6.52/8.19} &\makecell*[c]{8.87/10.02/\textbf{10.19}}         \\
			%\midrule
			\hline
			
			\multirow{2}{*}{Elec}     & HR@10      &\makecell*[c]{21.88/24.25/25.07}  &\makecell*[c]{17.37/19.22/20.18}  &\makecell*[c]{22.37/24.82/\textbf{25.19}}                    \\
			\multirow{2}{*}{}          & NDCG@10        &\makecell*[c]{11.37/14.22/15.18}&\makecell*[c]{8.80/9.28/10.82}&   \makecell*[c]{12.37/14.86/\textbf{15.11}}    \\
			\cline{2-5}
			%\specialrule{0em}{2pt}{2pt}
			% \midrule
			\hline
			\multirow{2}{*}{Phone}     & HR@10       & \makecell*[c]{27.37/27.22/27.99}           & \makecell*[c]{18.35/19.22/19.98} & \makecell*[c]{12.37/27.21/\textbf{28.82}}    \\
			\multirow{2}{*}{}          & NDCG@10        &   \makecell*[c]{16.07/16.92/17.19}     & \makecell*[c]{11.79/12.12/14.99}    &\makecell*[c]{12.96/17.21/\textbf{18.02}}    \\		
			%\bottomrule
			\hline
		\end{tabular}
	}
	\label{tb:pre_train_tasks}
\end{table}

\subsubsection{Effectiveness of the Pre-training Tasks (RQ2)} 
To explore whether the CL and ER tasks can benefit the Rec task, we propose three variant models, i.e., one performs only CL, one performs only ER, and one jointly performs both of them (denoted as A).
Besides, we examine the potential benefits of pre-training \sRC across domains by proposing three additional variant models, i.e.,  CL-, ER- and A-. We train these models using their respective loss functions only within domains.
Furthermore, we perform the original CL and ER task to explore whether unifying them as the MSL task could  benefit the Rec task. The corresponding variant models are CL*, ER* and A*. Table~\ref{tb:pre_train_tasks} shows the results~\footnote{In RQ2-RQ6, due to the space constraints, we only report the performance of the Dual-User-Intra CDR task. The remaining three tasks exhibit the same trend.}. We find that: 1) Jointly performing the CL and ER tasks leads to the best performance, suggesting that these two tasks can benefit the Rec task. 2) When performing each pre-training task separately,  CL is better than ER, as it focuses more on motif correlations. 
3)  Performing the pre-training tasks within each domain results in lower Rec performance, verifying that transferring knowledge across domains can benefit the Rec task. 
4) Compared with the original CL and ER tasks, unifying the CL and ER tasks as the MSL task leads to better performance, indicating that MSL can narrow the training gap between the pre-training and downstream tasks.

\subsubsection{Effectiveness of the Motif Types (RQ3)}
We report the Rec performance using different types of motifs in Table~\ref{tb:motifs}, where notations $\mathcal{B}$; $\mathcal{R}_3$, $\mathcal{R}_6$, $\mathcal{R}_9$; $\mathcal{T}_1$, $\mathcal{T}_2$, $\mathcal{T}_3$
denote the use of butterfly and random walks (with length of 3, 6, 9), and three different triangles as motifs.
We further integrate $\{ \mathcal{B}, \mathcal{R}_6, \mathcal{T} =  \{ \mathcal{T}_1, \mathcal{T}_2, \mathcal{T}_3 \}    \}$ as the motif, and denote this model variant as $\mathcal{A}$.
Following~\cite{conf/www/YuYLWH021}, we treat each type of motif as a channel, and use an attention mechanism~\cite{conf/www/YuYLWH021} to obtain fused node embeddings. The results show that: 1) Butterfly consistently yields the best performance in most cases, which implies capturing short-term dependencies of nodes with strong connections can benefit the knowledge transfer.
2) The performance first increases and then decreases as the random walk length becomes larger because shorter sequences lack useful information while longer sequences contain more noise.
3) Compared to $\mathcal{T}_1$, $\mathcal{T}_2$, $\mathcal{T}_3$,  combining all types of  triangles leads to the  performance gain, verifying that each triangle type is important. 4) In most cases, $\mathcal{A}$ does not perform best, as some motifs bring extra noise.

\begin{table}[t]
	\newcolumntype{?}{!{\vrule width 0.5pt}}
	\scriptsize
	\centering
	\caption{Performance comparison (\%) of motif types.}
	\setlength\tabcolsep{2.2pt}{
		\begin{tabular}{lc?cccccccccc}
			%\toprule
			\hline
			\multirow{2}{*}{Scenarios} & \multirow{2}{*}{Metrics}  &  \multicolumn{10}{c}{\makecell*[c]{Variant Models}}    \\  
			\cline{3-12}  & & \makecell*[c]{$\mathcal{B}$(\RC)} & $ \mathcal{R}_3$ & $\mathcal{R}_6$ & $\mathcal{R}_9$ &${\mathcal{T}_1}$ &${\mathcal{T}_2}$& ${\mathcal{T}_3}$ & $\mathcal{T}$&$\mathcal{A}$ &\\
			%\midrule
			\hline 
			
			\multirow{2}{*}{Sport}     & HR@10                                            &\makecell*[c]{19.23} & \makecell*[c]{11.03} &\makecell*[c]{16.91}&\makecell*[c]{16.66}&\makecell*[c]{10.23}&\makecell*[c]{9.11}&\makecell*[c]{12.18}&  \makecell*[c]{15.66}&{\makecell*[c]{\textbf{19.66}}}&\\
			\multirow{2}{*}{}          & NDCG@10                        & \makecell*[c]{\textbf{11.37}}    &6.37 &    8.12                   & 7.88&5.11&5.37&6.12&9.19&\makecell*[c]{11.13} & \\
			\cline{2-12}
			%\specialrule{0em}{2pt}{2pt}
			% \midrule
			\hline
			\multirow{2}{*}{Cloth}     & HR@10      & \makecell*[c]{\textbf{18.18}}& 11.92 &16.92   & 17.24 &10.23&12.35&11.24&17.23&\makecell*[c]{17.89}&\\
			\multirow{2}{*}{}          & NDCG@10                        & \makecell*[c]{\textbf{10.09}}        &8.34 &8.93       & 8.45   &4.56&4.67&6.32&8.88&  \makecell*[c]{8.40} &      \\
			%\midrule
			\hline
			
			\multirow{2}{*}{Elec}     & HR@10                          &\makecell*[c]{25.19}&20.12& \makecell*[c]{\textbf{25.24}}                        & 24.36 &20.35&21.35&20.56&24.89&\makecell*[c]{25.02}&\\
			\multirow{2}{*}{}          & NDCG@10        &\makecell*[c]{\textbf{15.11}}&9.31&14.54&14.31  &10.35&10.21&12.31&13.25&  \makecell*[c]{14.88} &          \\
			\cline{2-12}
			%\specialrule{0em}{2pt}{2pt}
			% \midrule
			\hline
			\multirow{2}{*}{Phone}     & HR@10                          &{\makecell*[c]{28.82}}     &19.35  &28.77& 26.24 &20.23&21.23&20.56&24.24&\makecell*[c]{\textbf{29.15}}&  \\
			\multirow{2}{*}{}          & NDCG@10 &\makecell*[c]{\textbf{18.02}} &12.35&17.35&16.31   &10.42&10.26&12.42&17.21&\makecell*[c]{17.98}&  \\		
			%\bottomrule
			\hline
		\end{tabular}
	}
	\label{tb:motifs}
\end{table}

\subsubsection{Effectiveness of the Encoder Components  (RQ4)}
We investigate the effect of  the hypergraph and MoDE Transformer by comparing \sRC with 7 variant models.
Notations ${\rm G}$, ${\rm H}$, ${\rm VT}$,  ${\rm MT}$  denote the use of LightGCN~\cite{conf/sigir/0001DWLZ020}, hypergraph, vanilla Transformer~\cite{conf/nips/VaswaniSPUJGKP17} and MoDE Transformer  as the base encoder, respectively; Notations ${\rm G}+ {\rm VT}$,   ${\rm G} + {\rm MT}$, ${\rm H} + {\rm VT}$ denote the use of  combining LightGCN and vanilla Transformer, LightGCN and MoDE Transformer, hypergraph and vanilla Transformer  as the base encoder, respectively. Aligning Table~\ref{tb:encoder} and Table~\ref{tb:dual_user_intra_inter}, we find that:  1) All  variant models exhibit competitive performance compared to the ``SOTA'' method. This suggests the choice of the encoder does not play a decisive role in the  Rec task.
2) Combining both the (hyper)graph and the (MoDE)Transformer as the base encoder leads to a performance gain, which indicates both these types of encoders help generate high-quality node embeddings. 3) Compared to the original Transformer, using a MoDE Transformer leads to a slight performance gain, which verifies MoDE architecture can model complex domain interactions. 4) Compared to LightGCN, performing hypergraph convolution leads to a slight performance gain, which verifies hypergraph can model complex nodes correlations.

\begin{table}[t]
	\newcolumntype{?}{!{\vrule width 0.5pt}}
	\scriptsize
	\centering
	\caption{Performance comparison (\%) of variant encoders.}
	\setlength\tabcolsep{2.25pt}{
		\begin{tabular}{lc?cccccccc}
			%\toprule
			\hline
			\multirow{2}{*}{Scenarios} & \multirow{2}{*}{Metrics} & \multicolumn{8}{c}{\makecell*[c]{Variant Models}}   \\  
			\cline{3-10}  & & \makecell*[c]{${\rm G}$} & ${\rm H}$ & ${\rm VT}$ & ${\rm MT}$ & ${\rm G + VT}$ &  ${\rm G +  MT}$ & ${\rm H + VT}$& ${\rm \RC}$   \\
			%\midrule
			\hline 
			
			\multirow{2}{*}{Sport}     & HR@10                & \makecell*[c]{18.77}&\makecell*[c]{18.88}& \makecell*[c]{18.81}       &\makecell*[c]{18.94}&\makecell*[c]{18.91} &\makecell*[c]{18.99} &\makecell*[c]{19.03} &\makecell*[c]{\textbf{19.23}} \\
			\multirow{2}{*}{}          & NDCG@10                        & \makecell*[c]{11.12}      &\makecell*[c]{11.13}&       \makecell*[c]{11.08}               & \makecell*[c]{11.11} & \makecell*[c]{11.24}& \makecell*[c]{11.30} &\makecell*[c]{11.31}&\makecell*[c]{\textbf{11.37}}\\
			\cline{2-10}
			%\specialrule{0em}{2pt}{2pt}
			% \midrule
			\hline
			\multirow{2}{*}{Cloth}     & HR@10                          &\makecell*[c]{17.91} &\makecell*[c]{17.93}& \makecell*[c]{17.88}                        & \makecell*[c]{18.02}& \makecell*[c]{17.93}&\makecell*[c]{18.01} &\makecell*[c]{18.03}&\makecell*[c]{\textbf{18.18}} \\
			\multirow{2}{*}{}          & NDCG@10                        &     \makecell*[c]{9.91}    &\makecell*[c]{9.92}&  \makecell*[c]{9.99}     &   \makecell*[c]{9.93}           & \makecell*[c]{10.01} &\makecell*[c]{10.07}   &\makecell*[c]{10.02} &\makecell*[c]{\textbf{10.19}}\\
			%\midrule
			\hline
			
			\multirow{2}{*}{Elec}     & HR@10                          & \makecell*[c]{24.99}&\makecell*[c]{24.94}&\makecell*[c]{24.98}                     & \makecell*[c]{25.13} & \makecell*[c]{25.01} & \makecell*[c]{25.09}&\makecell*[c]{25.04}& \makecell*[c]{\textbf{25.19}}\\
			\multirow{2}{*}{}          & NDCG@10        &\makecell*[c]{14.79}&\makecell*[c]{14.81}&\makecell*[c]{14.88}&     \makecell*[c]{14.93}           &          \makecell*[c]{14.98}         &     \makecell*[c]{15.08}   & \makecell*[c]{\textbf{15.19}}  & \makecell*[c]{15.11}   \\
			\cline{2-10}
			%\specialrule{0em}{2pt}{2pt}
			% \midrule
			\hline
			\multirow{2}{*}{Phone}     & HR@10                          &\makecell*[c]{28.44}                  & \makecell*[c]{28.48} &\makecell*[c]{28.52}&\makecell*[c]{28.53}&\makecell*[c]{28.71} &\makecell*[c]{28.68} &\makecell*[c]{28.77}& \makecell*[c]{\textbf{28.82}} \\
			\multirow{2}{*}{}          & NDCG@10                        &         \makecell*[c]{17.69}      &\makecell*[c]{17.75} &\makecell*[c]{17.71} &          \makecell*[c]{17.73}       &\makecell*[c]{17.91}  & \makecell*[c]{17.83} &\makecell*[c]{17.93} & \makecell*[c]{\textbf{18.02}} \\		
			%\bottomrule
			\hline
		\end{tabular}
	}
	\label{tb:encoder}
\end{table}

\begin{figure}[t]
	\centering
	\includegraphics[width= 0.435 \textwidth]{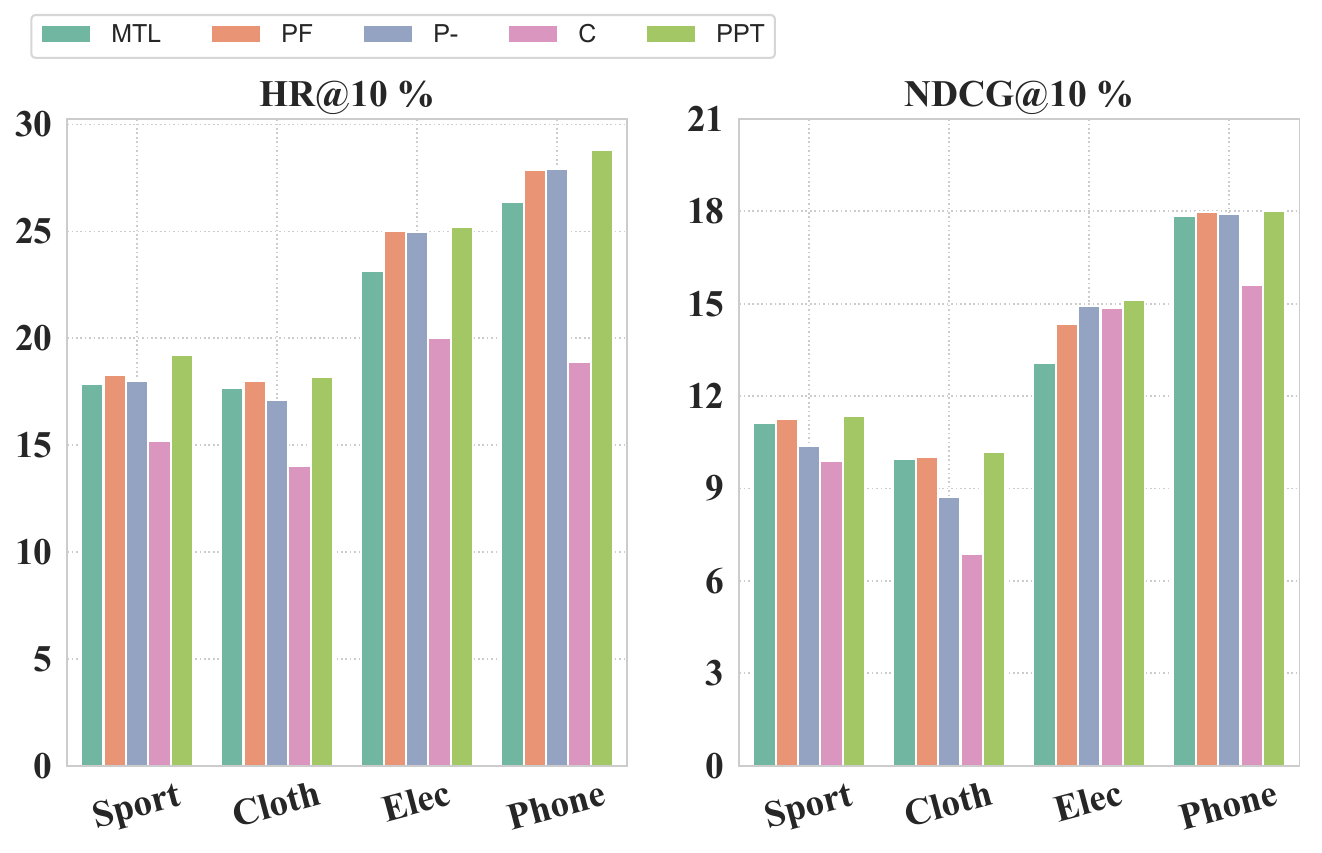}
	\caption{\label{fig:training_paradigm} Comparison of training paradigms. }
\end{figure}

\begin{figure}[t]
	\centering
	\includegraphics[width= 0.435 \textwidth]{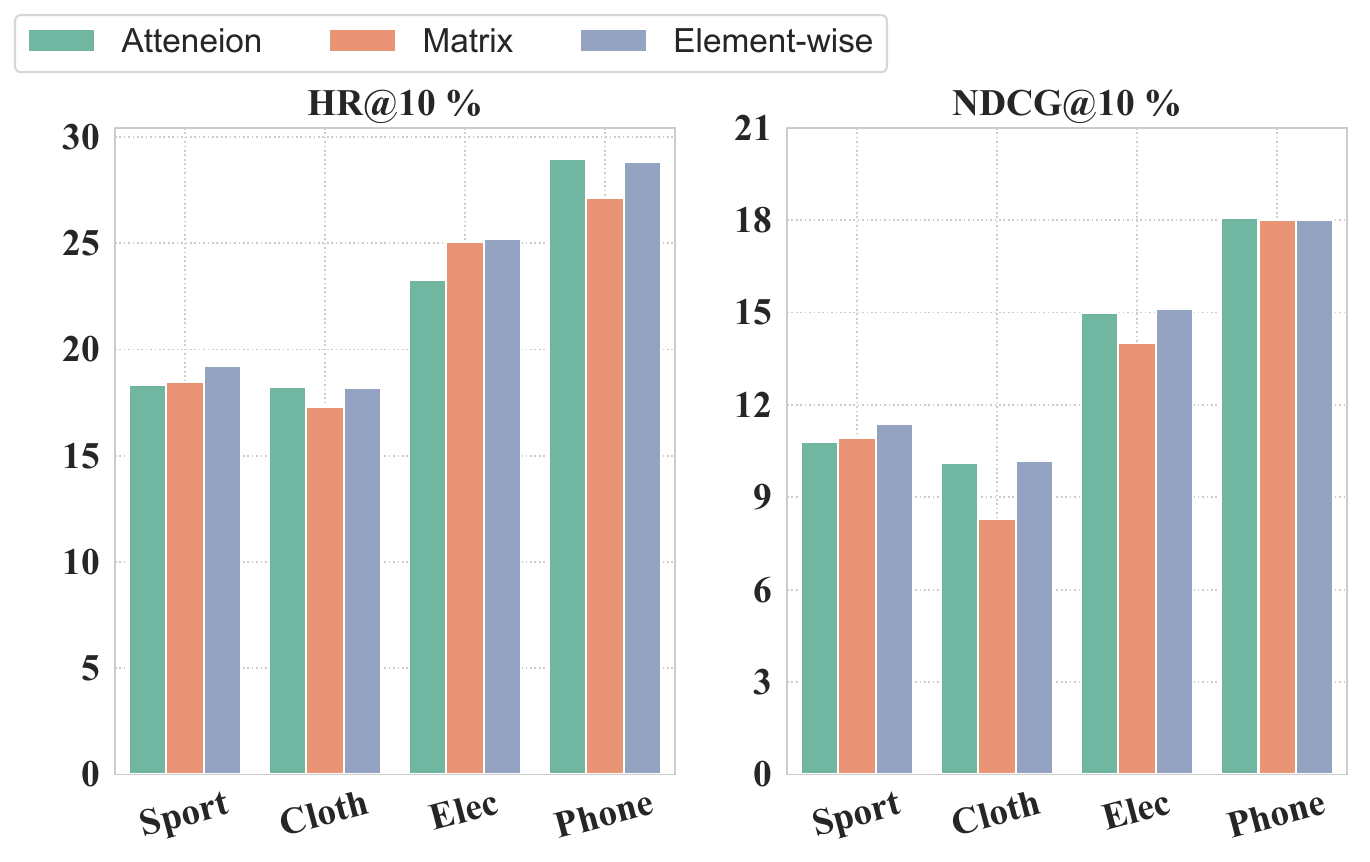}
	\caption{\label{fig:readout_function} Prompt-based  \textsc{ReadOut} operations comparison.}
\end{figure}

\subsubsection{Effectiveness of Prompt Tuning and Prompt-based $\textsc{ReadOut}$ Function (RQ5)}
We explore the benefits of prompt tuning by comparing \sRC with two training paradigms: MTL and PF.
In ${\rm PF}$, all parameters
are updated when we perform the Rec task; In ${\rm MTL}$, we discard the prompt tuning operation. 
Besides, we explore whether the prompt-based $\textsc{ReadOut}$ function is useful to summarize the motif by comparing \sRC with a variant model (denoted as ${\rm C}$). We discard the  $\textsc{ReadOut}$ function in this model and add the  CLS token upon each motif to obtain fused node embeddings.
We report the results in Fig.~\ref{fig:training_paradigm}.
We also compare the proposed three prompt tuning templates and report the  results in Fig.~\ref{fig:readout_function}. The results show that: 1) Compared to ``SOTA'' methods, ${\rm PPT}$ performs the best, verifying the superiority of the prompt tuning technique. 2) All these training paradigms exhibit competitive performance compared to ``SOTA'' methods, which implies the training paradigm is not a decisive factor, but the motif is a good choice to achieve general knowledge transfer. 3) ${\rm MTL}$ performs worse than ${\rm PF}$, as ${\rm PF}$ can reduce the training objective gap during the fine-tuning process. 
4) All of the $\textsc{ReadOut}$ functions exhibit competitive performance, as they can inspire the model's capability to handle the Rec task.
5) ${\rm C}$ performs worse than ${\rm PPT}$, which indicates only adding the CLS token is not sufficient for effectively summarizing motifs.

\begin{figure}[t]
	\centering
	\mbox{ 
		\subfigure[\scriptsize Sport]{\label{subfig:lambda_sport}
			\includegraphics[width=0.22\textwidth]{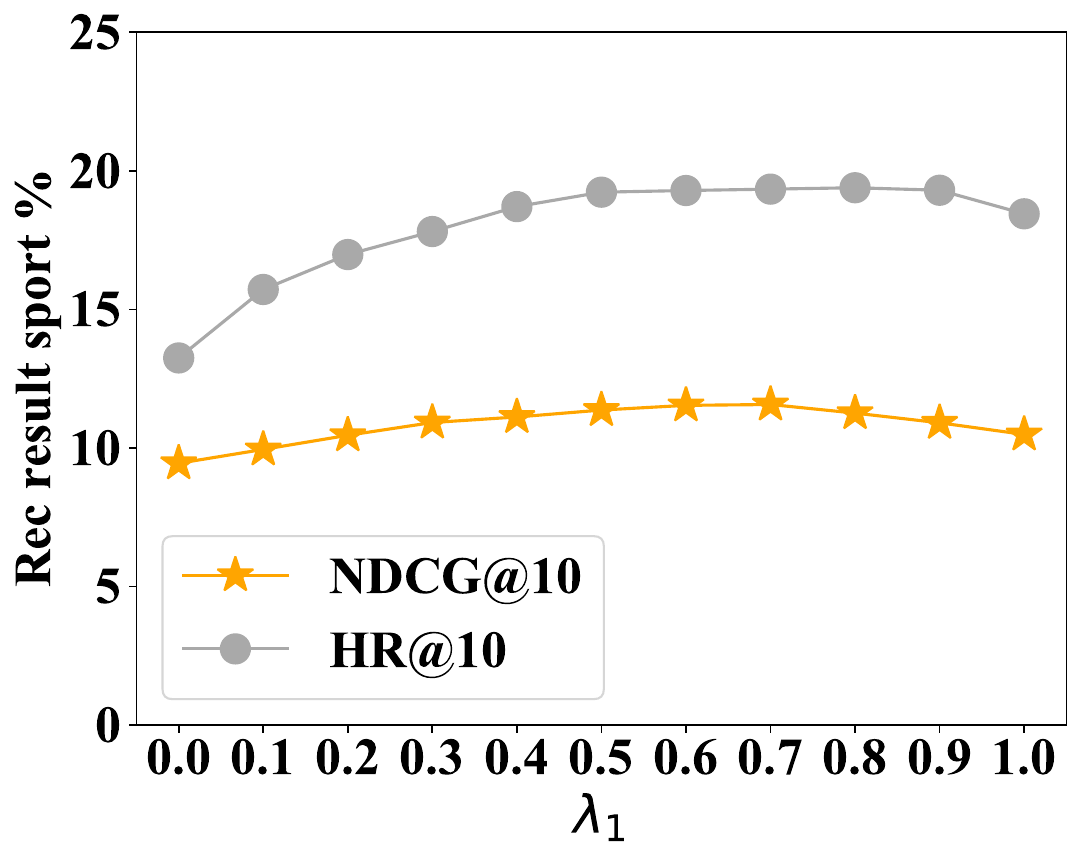}
		}
		
		\subfigure[\scriptsize Cloth]{\label{subfig:lambda_cloth}
			\includegraphics[width=0.22\textwidth]{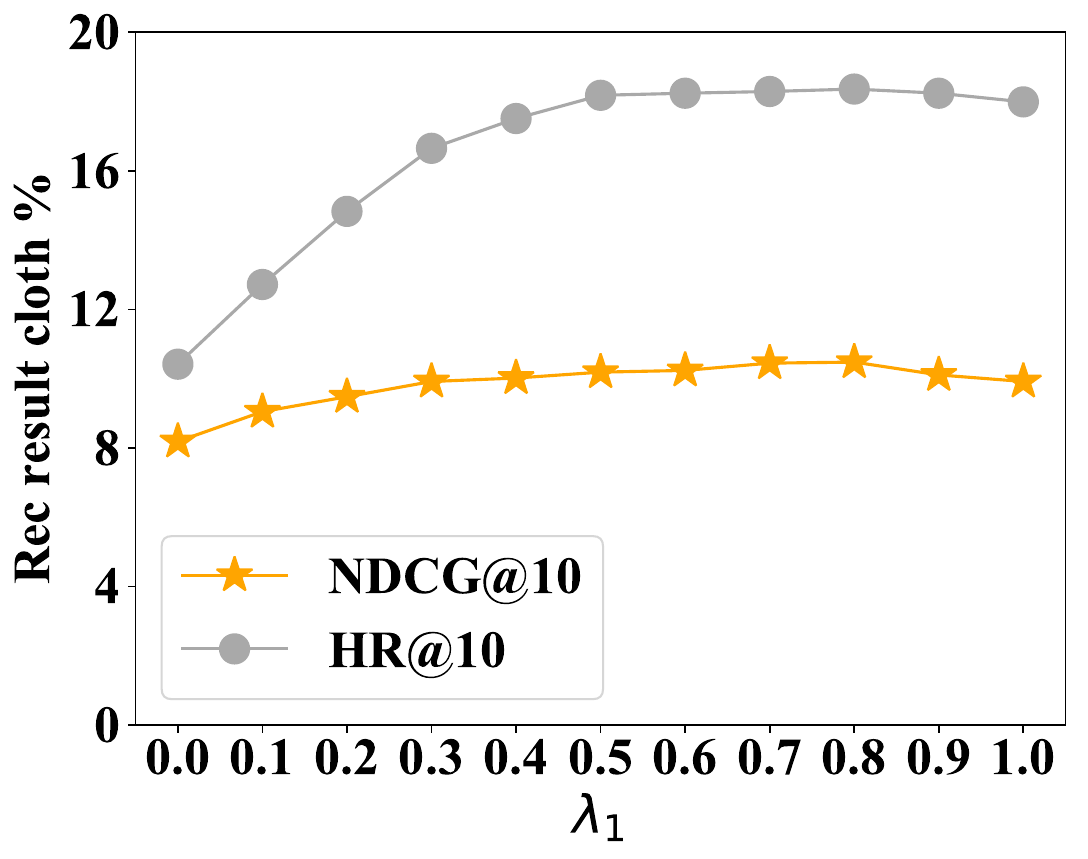}
		}

	}

	\hide{
	\mbox{ 
		\subfigure[\scriptsize Elec]{\label{subfig:lambda_elec}
		\includegraphics[width=0.20\textwidth]{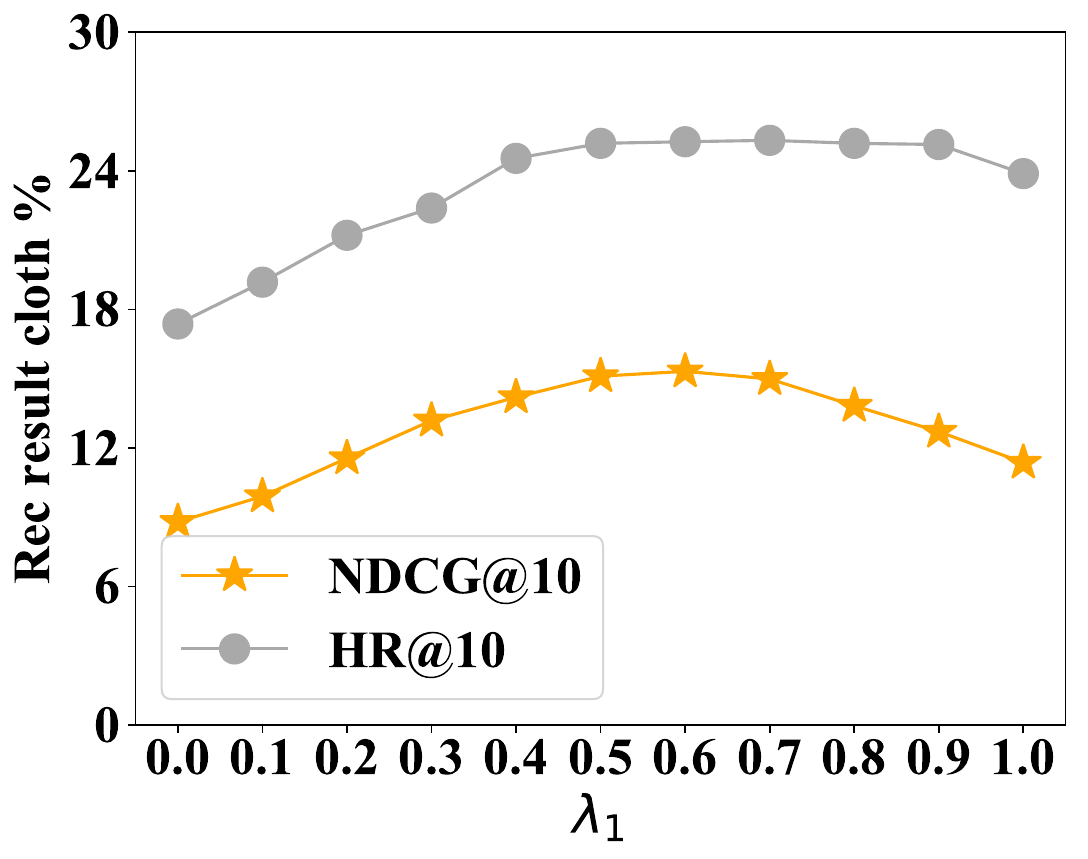}
		}
	
	\hspace{0.1cm}

		\subfigure[\scriptsize Phone]{\label{subfig:lambda_phone}
		\includegraphics[width=0.20\textwidth]{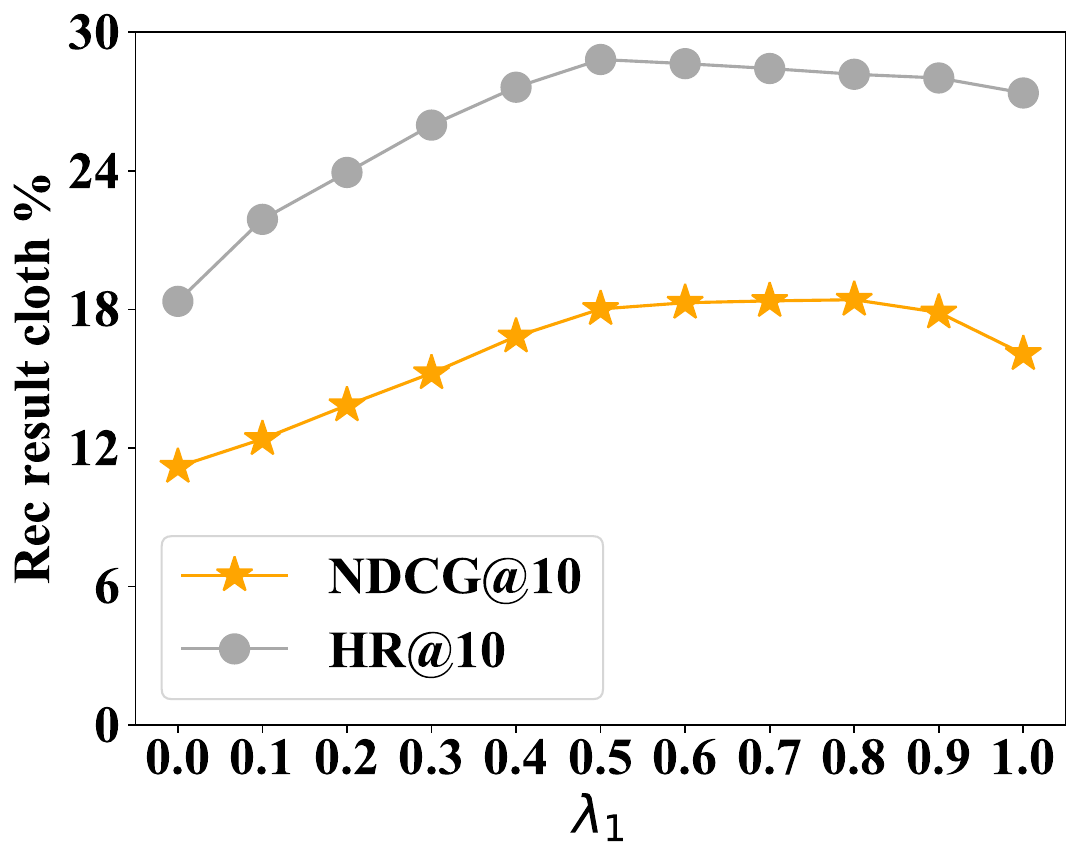}
		}
	}
}

	\caption{\label{fig:lambda} Impact of $\lambda_1$ on the Sport-Cloth dataset.  }
\end{figure}

\subsubsection{Effectiveness of the Hyperparameters (RQ6)} 
We examine the impact of the balance factor $\lambda_1$ and report the results in Fig.~\ref{fig:lambda}. We find that: 1) the Rec performance first increases, reaches a peak value of 7 in most cases, and then  declines. This suggests  CL  is more useful than ER, potentially due to its greater focus on motif correlations.
2) Performing only the CL ($\lambda_1$ = 1) or ER ($\lambda_1$ = 0) task results in poorer performance than performing them jointly, indicating that both tasks are necessary for the Rec task.

%% file: conclusion.tex
\section{Conclusion}
We propose \RC, which introduces  motifs to capture the general topology knowledge across domains in both intra-domain and inter-domain CDR tasks.  
Specifically, we devise three typical motifs: butterfly, triangle and random walk, and encode them through a Motif-based Encoder to obtain motif-based shared embeddings.
Additionally,  we train \sRC under the ``Pre-training \& Prompt Tuning''  paradigm to effectively transfer the general knowledge by unifying the pre-training and recommendation tasks as  a common motif-based similarity learning task, and introduce learnable prompt parameters to assist the model in
solving the downstream  recommendation task.
 Experimental results show the superiority of \sRC against the state-of-the-art methods on both intra-domain and inter-domain CDR tasks.

%% file: ack.tex
\section{ACKNOWLEDGMENTS}
This work is supported by National Natural Science Foundation of China(No.62172287, No.62102273), Australian Research Council under the streams of Future Fellowship (No. FT210100624) and Discovery Project (No. DP190101985).

%% file: sample-sigconf.bbl
%%% -*-BibTeX-*-
%%% Do NOT edit. File created by BibTeX with style
%%% ACM-Reference-Format-Journals [18-Jan-2012].

\begin{thebibliography}{57}

%%% ====================================================================
%%% NOTE TO THE USER: you can override these defaults by providing
%%% customized versions of any of these macros before the \bibliography
%%% command.  Each of them MUST provide its own final punctuation,
%%% except for \shownote{}, \showDOI{}, and \showURL{}.  The latter two
%%% do not use final punctuation, in order to avoid confusing it with
%%% the Web address.
%%%
%%% To suppress output of a particular field, define its macro to expand
%%% to an empty string, or better, \unskip, like this:
%%%
%%% \newcommand{\showDOI}[1]{\unskip}   % LaTeX syntax
%%%
%%% \def \showDOI #1{\unskip}           % plain TeX syntax
%%%
%%% ====================================================================

\ifx \showCODEN    \undefined \def \showCODEN     #1{\unskip}     \fi
\ifx \showDOI      \undefined \def \showDOI       #1{#1}\fi
\ifx \showISBNx    \undefined \def \showISBNx     #1{\unskip}     \fi
\ifx \showISBNxiii \undefined \def \showISBNxiii  #1{\unskip}     \fi
\ifx \showISSN     \undefined \def \showISSN      #1{\unskip}     \fi
\ifx \showLCCN     \undefined \def \showLCCN      #1{\unskip}     \fi
\ifx \shownote     \undefined \def \shownote      #1{#1}          \fi
\ifx \showarticletitle \undefined \def \showarticletitle #1{#1}   \fi
\ifx \showURL      \undefined \def \showURL       {\relax}        \fi
% The following commands are used for tagged output and should be
% invisible to TeX
\providecommand\bibfield[2]{#2}
\providecommand\bibinfo[2]{#2}
\providecommand\natexlab[1]{#1}
\providecommand\showeprint[2][]{arXiv:#2}

\bibitem[Bao et~al\mbox{.}(2022)]%
        {conf/nips/BaoW0LMASPW22}
\bibfield{author}{\bibinfo{person}{Hangbo Bao}, \bibinfo{person}{Wenhui Wang},
  \bibinfo{person}{Li Dong}, \bibinfo{person}{Qiang Liu},
  \bibinfo{person}{Owais~Khan Mohammed}, \bibinfo{person}{Kriti Aggarwal},
  \bibinfo{person}{Subhojit Som}, \bibinfo{person}{Songhao Piao}, {and}
  \bibinfo{person}{Furu Wei}.} \bibinfo{year}{2022}\natexlab{}.
\newblock \showarticletitle{VLMo: Unified Vision-Language Pre-Training with
  Mixture-of-Modality-Experts}. In \bibinfo{booktitle}{\emph{NeurIPS‘22}}.
\newblock


\bibitem[Benson et~al\mbox{.}(2016)]%
        {journals/science.aad9029}
\bibfield{author}{\bibinfo{person}{Austin~R. Benson}, \bibinfo{person}{David~F.
  Gleich}, {and} \bibinfo{person}{Jure Leskovec}.}
  \bibinfo{year}{2016}\natexlab{}.
\newblock \showarticletitle{Higher-order organization of complex networks}.
\newblock \bibinfo{journal}{\emph{Science}} \bibinfo{volume}{353},
  \bibinfo{number}{6295} (\bibinfo{year}{2016}), \bibinfo{pages}{163--166}.
\newblock


\bibitem[Bretto(2013)]%
        {bretto2013hypergraph}
\bibfield{author}{\bibinfo{person}{Alain Bretto}.}
  \bibinfo{year}{2013}\natexlab{}.
\newblock \showarticletitle{Hypergraph theory}.
\newblock \bibinfo{journal}{\emph{An introduction. Mathematical Engineering.
  Cham: Springer}} (\bibinfo{year}{2013}).
\newblock


\bibitem[Brown et~al\mbox{.}(2020)]%
        {brown2020language}
\bibfield{author}{\bibinfo{person}{Tom Brown}, \bibinfo{person}{Benjamin Mann},
  {et~al\mbox{.}}} \bibinfo{year}{2020}\natexlab{}.
\newblock \showarticletitle{Language models are few-shot learners}. In
  \bibinfo{booktitle}{\emph{NIPS'20}}, Vol.~\bibinfo{volume}{33}.
  \bibinfo{pages}{1877--1901}.
\newblock


\bibitem[Cao et~al\mbox{.}(2022a)]%
        {conf/sigir/CaoCLW22}
\bibfield{author}{\bibinfo{person}{Jiangxia Cao}, \bibinfo{person}{Xin Cong},
  \bibinfo{person}{Tingwen Liu}, {and} \bibinfo{person}{Bin Wang}.}
  \bibinfo{year}{2022}\natexlab{a}.
\newblock \showarticletitle{Item Similarity Mining for Multi-Market
  Recommendation}. In \bibinfo{booktitle}{\emph{SIGIR'22}}.
  \bibinfo{pages}{2249--2254}.
\newblock


\bibitem[Cao et~al\mbox{.}(2023)]%
        {conf/wsdm/CaoL0GL023}
\bibfield{author}{\bibinfo{person}{Jiangxia Cao}, \bibinfo{person}{Shaoshuai
  Li}, \bibinfo{person}{Bowen Yu}, \bibinfo{person}{Xiaobo Guo},
  \bibinfo{person}{Tingwen Liu}, {and} \bibinfo{person}{Bin Wang}.}
  \bibinfo{year}{2023}\natexlab{}.
\newblock \showarticletitle{Towards Universal Cross-Domain Recommendation}. In
  \bibinfo{booktitle}{\emph{WSDM'23}}. \bibinfo{pages}{78--86}.
\newblock


\bibitem[Cao et~al\mbox{.}(2022b)]%
        {conf/sigir/CaoLCYL022}
\bibfield{author}{\bibinfo{person}{Jiangxia Cao}, \bibinfo{person}{Xixun Lin},
  \bibinfo{person}{Xin Cong}, \bibinfo{person}{Jing Ya},
  \bibinfo{person}{Tingwen Liu}, {and} \bibinfo{person}{Bin Wang}.}
  \bibinfo{year}{2022}\natexlab{b}.
\newblock \showarticletitle{DisenCDR: Learning Disentangled Representations for
  Cross-Domain Recommendation}. In \bibinfo{booktitle}{\emph{SIGIR'22}}.
  \bibinfo{pages}{267--277}.
\newblock


\bibitem[Cao et~al\mbox{.}(2022c)]%
        {conf/icde/CaoSCLW22}
\bibfield{author}{\bibinfo{person}{Jiangxia Cao}, \bibinfo{person}{Jiawei
  Sheng}, \bibinfo{person}{Xin Cong}, \bibinfo{person}{Tingwen Liu}, {and}
  \bibinfo{person}{Bin Wang}.} \bibinfo{year}{2022}\natexlab{c}.
\newblock \showarticletitle{Cross-Domain Recommendation to Cold-Start Users via
  Variational Information Bottleneck}. In \bibinfo{booktitle}{\emph{ICDE'22}}.
  \bibinfo{pages}{2209--2223}.
\newblock


\bibitem[Chen et~al\mbox{.}(2020)]%
        {conf/icml/ChenK0H20}
\bibfield{author}{\bibinfo{person}{Ting Chen}, \bibinfo{person}{Simon
  Kornblith}, \bibinfo{person}{Mohammad Norouzi}, {and}
  \bibinfo{person}{Geoffrey~E. Hinton}.} \bibinfo{year}{2020}\natexlab{}.
\newblock \showarticletitle{A Simple Framework for Contrastive Learning of
  Visual Representations}. In \bibinfo{booktitle}{\emph{ICML'20}},
  Vol.~\bibinfo{volume}{119}. \bibinfo{pages}{1597--1607}.
\newblock


\bibitem[Gao et~al\mbox{.}(2021)]%
        {conf/acl/GaoFC20}
\bibfield{author}{\bibinfo{person}{Tianyu Gao}, \bibinfo{person}{Adam Fisch},
  {and} \bibinfo{person}{Danqi Chen}.} \bibinfo{year}{2021}\natexlab{}.
\newblock \showarticletitle{Making Pre-trained Language Models Better Few-shot
  Learners}. In \bibinfo{booktitle}{\emph{ACL/IJCNLP'21}}.
  \bibinfo{pages}{3816--3830}.
\newblock


\bibitem[Geng et~al\mbox{.}(2022)]%
        {conf/recsys/Geng0FGZ22}
\bibfield{author}{\bibinfo{person}{Shijie Geng}, \bibinfo{person}{Shuchang
  Liu}, \bibinfo{person}{Zuohui Fu}, \bibinfo{person}{Yingqiang Ge}, {and}
  \bibinfo{person}{Yongfeng Zhang}.} \bibinfo{year}{2022}\natexlab{}.
\newblock \showarticletitle{Recommendation as Language Processing {(RLP):} {A}
  Unified Pretrain, Personalized Prompt {\&} Predict Paradigm {(P5)}}. In
  \bibinfo{booktitle}{\emph{RecSys '22}}. \bibinfo{pages}{299--315}.
\newblock


\bibitem[Guo et~al\mbox{.}(2021)]%
        {conf/ijcai/0008T0ZNY21}
\bibfield{author}{\bibinfo{person}{Lei Guo}, \bibinfo{person}{Li Tang},
  \bibinfo{person}{Tong Chen}, \bibinfo{person}{Lei Zhu}, {et~al\mbox{.}}}
  \bibinfo{year}{2021}\natexlab{}.
\newblock \showarticletitle{{DA-GCN:} {A} Domain-aware Attentive Graph
  Convolution Network for Shared-account Cross-domain Sequential
  Recommendation}. In \bibinfo{booktitle}{\emph{IJCAI'21}}.
  \bibinfo{pages}{2483--2489}.
\newblock


\bibitem[Guo et~al\mbox{.}(2023)]%
        {journals/tkde/GuoZCWY23}
\bibfield{author}{\bibinfo{person}{Lei Guo}, \bibinfo{person}{Jinyu Zhang},
  \bibinfo{person}{Tong Chen}, \bibinfo{person}{Xinhua Wang}, {and}
  \bibinfo{person}{Hongzhi Yin}.} \bibinfo{year}{2023}\natexlab{}.
\newblock \showarticletitle{Reinforcement Learning-Enhanced Shared-Account
  Cross-Domain Sequential Recommendation}.
\newblock \bibinfo{journal}{\emph{{IEEE} Trans. Knowl. Data Eng.}}
  \bibinfo{volume}{35}, \bibinfo{number}{7} (\bibinfo{year}{2023}),
  \bibinfo{pages}{7397--7411}.
\newblock


\bibitem[Guo et~al\mbox{.}(2022)]%
        {guo2022time}
\bibfield{author}{\bibinfo{person}{Lei Guo}, \bibinfo{person}{Jinyu Zhang},
  \bibinfo{person}{Li Tang}, \bibinfo{person}{Tong Chen}, \bibinfo{person}{Lei
  Zhu}, {and} \bibinfo{person}{Hongzhi Yin}.} \bibinfo{year}{2022}\natexlab{}.
\newblock \showarticletitle{Time interval-enhanced graph neural network for
  shared-account cross-domain sequential recommendation}.
\newblock \bibinfo{journal}{\emph{IEEE Transactions on Neural Networks and
  Learning Systems}} (\bibinfo{year}{2022}).
\newblock


\bibitem[Gutmann and Hyv{\"{a}}rinen(2010)]%
        {journals/jmlr/GutmannH10}
\bibfield{author}{\bibinfo{person}{Michael Gutmann} {and} \bibinfo{person}{Aapo
  Hyv{\"{a}}rinen}.} \bibinfo{year}{2010}\natexlab{}.
\newblock \showarticletitle{Noise-contrastive estimation: {A} new estimation
  principle for unnormalized statistical models}. In
  \bibinfo{booktitle}{\emph{AISTATS'10}}.
\newblock


\bibitem[Hao et~al\mbox{.}(2023)]%
        {journals/tois/HaoYZLC23}
\bibfield{author}{\bibinfo{person}{Bowen Hao}, \bibinfo{person}{Hongzhi Yin},
  \bibinfo{person}{Jing Zhang}, \bibinfo{person}{Cuiping Li}, {and}
  \bibinfo{person}{Hong Chen}.} \bibinfo{year}{2023}\natexlab{}.
\newblock \showarticletitle{A Multi-strategy-based Pre-training Method for
  Cold-start Recommendation}.
\newblock \bibinfo{journal}{\emph{{ACM} Trans. Inf. Syst.}}
  \bibinfo{volume}{41}, \bibinfo{number}{2} (\bibinfo{year}{2023}),
  \bibinfo{pages}{31:1--31:24}.
\newblock


\bibitem[Hao et~al\mbox{.}(2021)]%
        {conf/wsdm/HaoZYL021}
\bibfield{author}{\bibinfo{person}{Bowen Hao}, \bibinfo{person}{Jing Zhang},
  \bibinfo{person}{Hongzhi Yin}, \bibinfo{person}{Cuiping Li}, {and}
  \bibinfo{person}{Hong Chen}.} \bibinfo{year}{2021}\natexlab{}.
\newblock \showarticletitle{Pre-Training Graph Neural Networks for Cold-Start
  Users and Items Representation}. In \bibinfo{booktitle}{\emph{WSDM'21}}.
  \bibinfo{pages}{265--273}.
\newblock


\bibitem[He et~al\mbox{.}(2020)]%
        {conf/sigir/0001DWLZ020}
\bibfield{author}{\bibinfo{person}{Xiangnan He}, \bibinfo{person}{Kuan Deng},
  \bibinfo{person}{Xiang Wang}, \bibinfo{person}{Yan Li},
  \bibinfo{person}{Yong{-}Dong Zhang}, {and} \bibinfo{person}{Meng Wang}.}
  \bibinfo{year}{2020}\natexlab{}.
\newblock \showarticletitle{LightGCN: Simplifying and Powering Graph
  Convolution Network for Recommendation}. In
  \bibinfo{booktitle}{\emph{SIGIR'20}}. \bibinfo{pages}{639--648}.
\newblock


\bibitem[Hu et~al\mbox{.}(2018)]%
        {conf/cikm/HuZY18}
\bibfield{author}{\bibinfo{person}{Guangneng Hu}, \bibinfo{person}{Yu Zhang},
  {and} \bibinfo{person}{Qiang Yang}.} \bibinfo{year}{2018}\natexlab{}.
\newblock \showarticletitle{CoNet: Collaborative Cross Networks for
  Cross-Domain Recommendation}. In \bibinfo{booktitle}{\emph{CIKM'18}}.
  \bibinfo{pages}{667--676}.
\newblock


\bibitem[IV et~al\mbox{.}(2022)]%
        {conf/acl/LoganBWP0022}
\bibfield{author}{\bibinfo{person}{Robert L.~Logan IV}, \bibinfo{person}{Ivana
  Balazevic}, \bibinfo{person}{Eric Wallace}, \bibinfo{person}{Fabio Petroni},
  \bibinfo{person}{Sameer Singh}, {and} \bibinfo{person}{Sebastian Riedel}.}
  \bibinfo{year}{2022}\natexlab{}.
\newblock \showarticletitle{Cutting Down on Prompts and Parameters: Simple
  Few-Shot Learning with Language Models}. In
  \bibinfo{booktitle}{\emph{Findings of the ACL'22}}.
  \bibinfo{pages}{2824--2835}.
\newblock


\bibitem[Kanagawa et~al\mbox{.}(2019)]%
        {conf/ecir/KanagawaKSTS19}
\bibfield{author}{\bibinfo{person}{Heishiro Kanagawa}, \bibinfo{person}{Hayato
  Kobayashi}, \bibinfo{person}{Nobuyuki Shimizu}, \bibinfo{person}{Yukihiro
  Tagami}, {and} \bibinfo{person}{Taiji Suzuki}.}
  \bibinfo{year}{2019}\natexlab{}.
\newblock \showarticletitle{Cross-Domain Recommendation via Deep Domain
  Adaptation}. In \bibinfo{booktitle}{\emph{ECIR'19}},
  Vol.~\bibinfo{volume}{11438}. \bibinfo{pages}{20--29}.
\newblock


\bibitem[Kang et~al\mbox{.}(2019)]%
        {conf/cikm/KangHLY19}
\bibfield{author}{\bibinfo{person}{SeongKu Kang}, \bibinfo{person}{Junyoung
  Hwang}, \bibinfo{person}{Dongha Lee}, {and} \bibinfo{person}{Hwanjo Yu}.}
  \bibinfo{year}{2019}\natexlab{}.
\newblock \showarticletitle{Semi-Supervised Learning for Cross-Domain
  Recommendation to Cold-Start Users}. In \bibinfo{booktitle}{\emph{CIKM'19}}.
  \bibinfo{pages}{1563--1572}.
\newblock


\bibitem[Krichene and Rendle(2020)]%
        {conf/kdd/KricheneR20}
\bibfield{author}{\bibinfo{person}{Walid Krichene} {and}
  \bibinfo{person}{Steffen Rendle}.} \bibinfo{year}{2020}\natexlab{}.
\newblock \showarticletitle{On Sampled Metrics for Item Recommendation}. In
  \bibinfo{booktitle}{\emph{KDD'20}}. \bibinfo{pages}{1748--1757}.
\newblock


\bibitem[Li et~al\mbox{.}(2023a)]%
        {conf/wsdm/LiXYHLSQN23}
\bibfield{author}{\bibinfo{person}{Chenglin Li}, \bibinfo{person}{Yuanzhen
  Xie}, \bibinfo{person}{Chenyun Yu}, \bibinfo{person}{Bo Hu},
  \bibinfo{person}{Zang Li}, \bibinfo{person}{Guoqiang Shu},
  \bibinfo{person}{Xiaohu Qie}, {and} \bibinfo{person}{Di Niu}.}
  \bibinfo{year}{2023}\natexlab{a}.
\newblock \showarticletitle{One for All, All for One: Learning and Transferring
  User Embeddings for Cross-Domain Recommendation}. In
  \bibinfo{booktitle}{\emph{WSDM'23}}. \bibinfo{pages}{366--374}.
\newblock


\bibitem[Li et~al\mbox{.}(2023b)]%
        {journals/corr/abs-2202-07371}
\bibfield{author}{\bibinfo{person}{Lei Li}, \bibinfo{person}{Yongfeng Zhang},
  {and} \bibinfo{person}{Li Chen}.} \bibinfo{year}{2023}\natexlab{b}.
\newblock \showarticletitle{Personalized Prompt Learning for Explainable
  Recommendation}.
\newblock \bibinfo{journal}{\emph{{ACM} Trans. Inf. Syst.}}
  \bibinfo{volume}{abs/2202.07371} (\bibinfo{year}{2023}).
\newblock


\bibitem[Lian et~al\mbox{.}(2017)]%
        {conf/www/LianZXS17}
\bibfield{author}{\bibinfo{person}{Jianxun Lian}, \bibinfo{person}{Fuzheng
  Zhang}, \bibinfo{person}{Xing Xie}, {and} \bibinfo{person}{Guangzhong Sun}.}
  \bibinfo{year}{2017}\natexlab{}.
\newblock \showarticletitle{CCCFNet: {A} Content-Boosted Collaborative
  Filtering Neural Network for Cross Domain Recommender Systems}. In
  \bibinfo{booktitle}{\emph{WWW'17}}. \bibinfo{pages}{817--818}.
\newblock


\bibitem[Liu et~al\mbox{.}(2020b)]%
        {conf/www/LiuZZLSX0X20}
\bibfield{author}{\bibinfo{person}{Jian Liu}, \bibinfo{person}{Pengpeng Zhao},
  \bibinfo{person}{Fuzhen Zhuang}, \bibinfo{person}{Yanchi Liu},
  \bibinfo{person}{Victor~S. Sheng}, \bibinfo{person}{Jiajie Xu},
  \bibinfo{person}{Xiaofang Zhou}, {and} \bibinfo{person}{Hui Xiong}.}
  \bibinfo{year}{2020}\natexlab{b}.
\newblock \showarticletitle{Exploiting Aesthetic Preference in Deep Cross
  Networks for Cross-domain Recommendation}. In
  \bibinfo{booktitle}{\emph{WWW'20}}. \bibinfo{pages}{2768--2774}.
\newblock


\bibitem[Liu et~al\mbox{.}(2020a)]%
        {conf/cikm/LiuLLP20}
\bibfield{author}{\bibinfo{person}{Meng Liu}, \bibinfo{person}{Jianjun Li},
  \bibinfo{person}{Guohui Li}, {and} \bibinfo{person}{Peng Pan}.}
  \bibinfo{year}{2020}\natexlab{a}.
\newblock \showarticletitle{Cross Domain Recommendation via Bi-directional
  Transfer Graph Collaborative Filtering Networks}. In
  \bibinfo{booktitle}{\emph{CIKM'20}}. \bibinfo{pages}{885--894}.
\newblock


\bibitem[Liu et~al\mbox{.}(2022)]%
        {conf/www/LiuZH022}
\bibfield{author}{\bibinfo{person}{Weiming Liu}, \bibinfo{person}{Xiaolin
  Zheng}, \bibinfo{person}{Mengling Hu}, {and} \bibinfo{person}{Chaochao
  Chen}.} \bibinfo{year}{2022}\natexlab{}.
\newblock \showarticletitle{Collaborative Filtering with Attribution Alignment
  for Review-based Non-overlapped Cross Domain Recommendation}. In
  \bibinfo{booktitle}{\emph{WWW'22}}. \bibinfo{pages}{1181--1190}.
\newblock


\bibitem[Liu et~al\mbox{.}(2021)]%
        {journals/corr/abs-2103-10385}
\bibfield{author}{\bibinfo{person}{Xiao Liu}, \bibinfo{person}{Yanan Zheng},
  \bibinfo{person}{Zhengxiao Du}, \bibinfo{person}{Ming Ding},
  \bibinfo{person}{Yujie Qian}, \bibinfo{person}{Zhilin Yang}, {and}
  \bibinfo{person}{Jie Tang}.} \bibinfo{year}{2021}\natexlab{}.
\newblock \showarticletitle{{GPT} Understands, Too}.
\newblock \bibinfo{journal}{\emph{CoRR}}  \bibinfo{volume}{abs/2103.10385}
  (\bibinfo{year}{2021}).
\newblock


\bibitem[Liu et~al\mbox{.}(2023)]%
        {conf/www/LiuY0023}
\bibfield{author}{\bibinfo{person}{Zemin Liu}, \bibinfo{person}{Xingtong Yu},
  \bibinfo{person}{Yuan Fang}, {and} \bibinfo{person}{Xinming Zhang}.}
  \bibinfo{year}{2023}\natexlab{}.
\newblock \showarticletitle{GraphPrompt: Unifying Pre-Training and Downstream
  Tasks for Graph Neural Networks}. In \bibinfo{booktitle}{\emph{WWW'23}}.
  \bibinfo{pages}{417--428}.
\newblock


\bibitem[Loughry et~al\mbox{.}(2002)]%
        {journals/JJL02}
\bibfield{author}{\bibinfo{person}{J. Loughry}, \bibinfo{person}{J.I. van
  Hemert}, {and} \bibinfo{person}{L. Schoofs}.}
  \bibinfo{year}{2002}\natexlab{}.
\newblock \showarticletitle{Efficiently Enumerating the Subsets of a Set}.
\newblock \bibinfo{journal}{\emph{applied math}} (\bibinfo{year}{2002}).
\newblock


\bibitem[Milo et~al\mbox{.}(2002)]%
        {milo2002network}
\bibfield{author}{\bibinfo{person}{Ron Milo}, \bibinfo{person}{Shai Shen-Orr},
  \bibinfo{person}{Shalev Itzkovitz}, \bibinfo{person}{Nadav Kashtan},
  \bibinfo{person}{Dmitri Chklovskii}, {and} \bibinfo{person}{Uri Alon}.}
  \bibinfo{year}{2002}\natexlab{}.
\newblock \showarticletitle{Network motifs: simple building blocks of complex
  networks}.
\newblock \bibinfo{journal}{\emph{Science}} \bibinfo{volume}{298},
  \bibinfo{number}{5594} (\bibinfo{year}{2002}), \bibinfo{pages}{824--827}.
\newblock


\bibitem[Moreno et~al\mbox{.}(2012)]%
        {conf/cikm/MorenoSRS12}
\bibfield{author}{\bibinfo{person}{Orly Moreno}, \bibinfo{person}{Bracha
  Shapira}, \bibinfo{person}{Lior Rokach}, {and} \bibinfo{person}{Guy Shani}.}
  \bibinfo{year}{2012}\natexlab{}.
\newblock \showarticletitle{{TALMUD:} transfer learning for multiple domains}.
  In \bibinfo{booktitle}{\emph{CIKM'12}}. \bibinfo{pages}{425--434}.
\newblock


\bibitem[Ouyang et~al\mbox{.}(2022)]%
        {journals/corr/abs-2203-02155}
\bibfield{author}{\bibinfo{person}{Long Ouyang}, \bibinfo{person}{Jeff Wu},
  \bibinfo{person}{Xu Jiang}, {and} \bibinfo{person}{et al.}}
  \bibinfo{year}{2022}\natexlab{}.
\newblock \showarticletitle{Training language models to follow instructions
  with human feedback}.
\newblock \bibinfo{journal}{\emph{CoRR}}  \bibinfo{volume}{abs/2203.02155}
  (\bibinfo{year}{2022}).
\newblock


\bibitem[Perozzi et~al\mbox{.}(2014)]%
        {conf/kdd/PerozziAS14}
\bibfield{author}{\bibinfo{person}{Bryan Perozzi}, \bibinfo{person}{Rami
  Al{-}Rfou}, {and} \bibinfo{person}{Steven Skiena}.}
  \bibinfo{year}{2014}\natexlab{}.
\newblock \showarticletitle{DeepWalk: online learning of social
  representations}. In \bibinfo{booktitle}{\emph{KDD'14}}.
  \bibinfo{pages}{701--710}.
\newblock


\bibitem[Rafailidis and Crestani(2017)]%
        {conf/cikm/RafailidisC17}
\bibfield{author}{\bibinfo{person}{Dimitrios Rafailidis} {and}
  \bibinfo{person}{Fabio Crestani}.} \bibinfo{year}{2017}\natexlab{}.
\newblock \showarticletitle{A Collaborative Ranking Model for Cross-Domain
  Recommendations}. In \bibinfo{booktitle}{\emph{CIKM'17}}.
  \bibinfo{pages}{2263--2266}.
\newblock


\bibitem[Sanei{-}Mehri et~al\mbox{.}(2018)]%
        {conf/kdd/Sanei-MehriST18}
\bibfield{author}{\bibinfo{person}{Seyed{-}Vahid Sanei{-}Mehri},
  \bibinfo{person}{Ahmet~Erdem Sariy{\"{u}}ce}, {and} \bibinfo{person}{Srikanta
  Tirthapura}.} \bibinfo{year}{2018}\natexlab{}.
\newblock \showarticletitle{Butterfly Counting in Bipartite Networks}. In
  \bibinfo{booktitle}{\emph{KDD'18}}. \bibinfo{pages}{2150--2159}.
\newblock


\bibitem[Shin et~al\mbox{.}(2020)]%
        {conf/emnlp/ShinRLWS20}
\bibfield{author}{\bibinfo{person}{Taylor Shin}, \bibinfo{person}{Yasaman
  Razeghi}, \bibinfo{person}{Robert L.~Logan IV}, \bibinfo{person}{Eric
  Wallace}, {and} \bibinfo{person}{Sameer Singh}.}
  \bibinfo{year}{2020}\natexlab{}.
\newblock \showarticletitle{AutoPrompt: Eliciting Knowledge from Language
  Models with Automatically Generated Prompts}. In
  \bibinfo{booktitle}{\emph{EMNLP‘20}}. \bibinfo{pages}{4222--4235}.
\newblock


\bibitem[Sileo et~al\mbox{.}(2022)]%
        {conf/ecir/SileoVR22}
\bibfield{author}{\bibinfo{person}{Damien Sileo}, \bibinfo{person}{Wout
  Vossen}, {and} \bibinfo{person}{Robbe Raymaekers}.}
  \bibinfo{year}{2022}\natexlab{}.
\newblock \showarticletitle{Zero-Shot Recommendation as Language Modeling}. In
  \bibinfo{booktitle}{\emph{ECIR'22}} \emph{(\bibinfo{series}{Lecture Notes in
  Computer Science}, Vol.~\bibinfo{volume}{13186})}. \bibinfo{pages}{223--230}.
\newblock


\bibitem[Steck(2019)]%
        {conf/www/Steck19}
\bibfield{author}{\bibinfo{person}{Harald Steck}.}
  \bibinfo{year}{2019}\natexlab{}.
\newblock \showarticletitle{Embarrassingly Shallow Autoencoders for Sparse
  Data}. In \bibinfo{booktitle}{\emph{WWW'19}}. \bibinfo{pages}{3251--3257}.
\newblock


\bibitem[Sun et~al\mbox{.}(2022)]%
        {conf/kdd/SunZHWW22}
\bibfield{author}{\bibinfo{person}{Mingchen Sun}, \bibinfo{person}{Kaixiong
  Zhou}, \bibinfo{person}{Xin He}, \bibinfo{person}{Ying Wang}, {and}
  \bibinfo{person}{Xin Wang}.} \bibinfo{year}{2022}\natexlab{}.
\newblock \showarticletitle{{GPPT:} Graph Pre-training and Prompt Tuning to
  Generalize Graph Neural Networks}. In \bibinfo{booktitle}{\emph{KDD'22}}.
  \bibinfo{pages}{1717--1727}.
\newblock


\bibitem[Vaswani et~al\mbox{.}(2017)]%
        {conf/nips/VaswaniSPUJGKP17}
\bibfield{author}{\bibinfo{person}{Ashish Vaswani}, \bibinfo{person}{Noam
  Shazeer}, \bibinfo{person}{Niki Parmar}, \bibinfo{person}{Jakob Uszkoreit},
  \bibinfo{person}{Llion Jones}, \bibinfo{person}{Aidan~N. Gomez},
  \bibinfo{person}{Lukasz Kaiser}, {and} \bibinfo{person}{Illia Polosukhin}.}
  \bibinfo{year}{2017}\natexlab{}.
\newblock \showarticletitle{Attention is All you Need}. In
  \bibinfo{booktitle}{\emph{NIPS'17}}.
\newblock


\bibitem[Wang et~al\mbox{.}(2019)]%
        {journals/pvldb/WangLQZZ19}
\bibfield{author}{\bibinfo{person}{Kai Wang}, \bibinfo{person}{Xuemin Lin},
  \bibinfo{person}{Lu Qin}, \bibinfo{person}{Wenjie Zhang}, {and}
  \bibinfo{person}{Ying Zhang}.} \bibinfo{year}{2019}\natexlab{}.
\newblock \showarticletitle{Vertex Priority Based Butterfly Counting for
  Large-scale Bipartite Networks}.
\newblock \bibinfo{journal}{\emph{VLDB'19}} (\bibinfo{year}{2019}),
  \bibinfo{pages}{1139--1152}.
\newblock


\bibitem[Wu et~al\mbox{.}(2022)]%
        {conf/sigir/WuXZZ0ZL022}
\bibfield{author}{\bibinfo{person}{Yiqing Wu}, \bibinfo{person}{Ruobing Xie},
  \bibinfo{person}{Yongchun Zhu}, \bibinfo{person}{Fuzhen Zhuang},
  \bibinfo{person}{Xiang Ao}, \bibinfo{person}{Xu Zhang}, \bibinfo{person}{Leyu
  Lin}, {and} \bibinfo{person}{Qing He}.} \bibinfo{year}{2022}\natexlab{}.
\newblock \showarticletitle{Selective Fairness in Recommendation via Prompts}.
  In \bibinfo{booktitle}{\emph{SIGIR '22}}. \bibinfo{pages}{2657--2662}.
\newblock


\bibitem[Yang et~al\mbox{.}(2017)]%
        {conf/sigir/YangYYLC17}
\bibfield{author}{\bibinfo{person}{Chunfeng Yang}, \bibinfo{person}{Huan Yan},
  \bibinfo{person}{Donghan Yu}, \bibinfo{person}{Yong Li}, {and}
  \bibinfo{person}{Dah~Ming Chiu}.} \bibinfo{year}{2017}\natexlab{}.
\newblock \showarticletitle{Multi-site User Behavior Modeling and Its
  Application in Video Recommendation}. In
  \bibinfo{booktitle}{\emph{SIGIR'17}}. \bibinfo{pages}{175--184}.
\newblock


\bibitem[Yin et~al\mbox{.}(2019)]%
        {conf/icde/YinW0LYZ19}
\bibfield{author}{\bibinfo{person}{Hongzhi Yin}, \bibinfo{person}{Qinyong
  Wang}, \bibinfo{person}{Kai Zheng}, \bibinfo{person}{Zhixu Li},
  \bibinfo{person}{Jiali Yang}, {and} \bibinfo{person}{Xiaofang Zhou}.}
  \bibinfo{year}{2019}\natexlab{}.
\newblock \showarticletitle{Social Influence-Based Group Representation
  Learning for Group Recommendation}. In \bibinfo{booktitle}{\emph{ICDE'19}}.
  \bibinfo{publisher}{{IEEE}}, \bibinfo{pages}{566--577}.
\newblock


\bibitem[Yu et~al\mbox{.}(2021a)]%
        {conf/kdd/YuY000H21}
\bibfield{author}{\bibinfo{person}{Junliang Yu}, \bibinfo{person}{Hongzhi Yin},
  \bibinfo{person}{Min Gao}, \bibinfo{person}{Xin Xia},
  \bibinfo{person}{Xiangliang Zhang}, {and} \bibinfo{person}{Nguyen Quoc~Viet
  Hung}.} \bibinfo{year}{2021}\natexlab{a}.
\newblock \showarticletitle{Socially-Aware Self-Supervised Tri-Training for
  Recommendation}. In \bibinfo{booktitle}{\emph{KDD'21}}.
  \bibinfo{publisher}{{ACM}}, \bibinfo{pages}{2084--2092}.
\newblock


\bibitem[Yu et~al\mbox{.}(2021b)]%
        {conf/www/YuYLWH021}
\bibfield{author}{\bibinfo{person}{Junliang Yu}, \bibinfo{person}{Hongzhi Yin},
  \bibinfo{person}{Jundong Li}, \bibinfo{person}{Qinyong Wang},
  \bibinfo{person}{Nguyen Quoc~Viet Hung}, {and} \bibinfo{person}{Xiangliang
  Zhang}.} \bibinfo{year}{2021}\natexlab{b}.
\newblock \showarticletitle{Self-Supervised Multi-Channel Hypergraph
  Convolutional Network for Social Recommendation}. In
  \bibinfo{booktitle}{\emph{WWW'21}}. \bibinfo{pages}{413--424}.
\newblock


\bibitem[Yu et~al\mbox{.}(2023)]%
        {journals/corr/abs-2203-15876}
\bibfield{author}{\bibinfo{person}{Junliang Yu}, \bibinfo{person}{Hongzhi Yin},
  \bibinfo{person}{Xin Xia}, \bibinfo{person}{Tong Chen},
  \bibinfo{person}{Jundong Li}, {and} \bibinfo{person}{Zi Huang}.}
  \bibinfo{year}{2023}\natexlab{}.
\newblock \showarticletitle{Self-Supervised Learning for Recommender Systems:
  {A} Survey}.
\newblock \bibinfo{journal}{\emph{{IEEE} Trans. Knowl. Data Eng.}}
  (\bibinfo{year}{2023}).
\newblock


\bibitem[Yuan et~al\mbox{.}(2019)]%
        {conf/ijcai/YuanYB19}
\bibfield{author}{\bibinfo{person}{Feng Yuan}, \bibinfo{person}{Lina Yao},
  {and} \bibinfo{person}{Boualem Benatallah}.} \bibinfo{year}{2019}\natexlab{}.
\newblock \showarticletitle{DARec: Deep Domain Adaptation for Cross-Domain
  Recommendation via Transferring Rating Patterns}. In
  \bibinfo{booktitle}{\emph{IJCAI‘19}}. \bibinfo{pages}{4227--4233}.
\newblock


\bibitem[Zhang and Wang(2023)]%
        {conf/sigir/ZhangW23}
\bibfield{author}{\bibinfo{person}{Zizhuo Zhang} {and} \bibinfo{person}{Bang
  Wang}.} \bibinfo{year}{2023}\natexlab{}.
\newblock \showarticletitle{Prompt Learning for News Recommendation}. In
  \bibinfo{booktitle}{\emph{SIGIR'23}}. \bibinfo{pages}{227--237}.
\newblock


\bibitem[Zhao et~al\mbox{.}(2013)]%
        {conf/aaai/ZhaoPXZLY13}
\bibfield{author}{\bibinfo{person}{Lili Zhao}, \bibinfo{person}{Sinno~Jialin
  Pan}, \bibinfo{person}{Evan~Wei Xiang}, \bibinfo{person}{Erheng Zhong},
  \bibinfo{person}{Zhongqi Lu}, {and} \bibinfo{person}{Qiang Yang}.}
  \bibinfo{year}{2013}\natexlab{}.
\newblock \showarticletitle{Active Transfer Learning for Cross-System
  Recommendation}. In \bibinfo{booktitle}{\emph{AAAI'13}}.
\newblock


\bibitem[Zhou et~al\mbox{.}(2020)]%
        {conf/cikm/ZhouWZZWZWW20}
\bibfield{author}{\bibinfo{person}{Kun Zhou}, \bibinfo{person}{Hui Wang},
  \bibinfo{person}{Wayne~Xin Zhao}, \bibinfo{person}{Yutao Zhu},
  \bibinfo{person}{Sirui Wang}, \bibinfo{person}{Fuzheng Zhang},
  \bibinfo{person}{Zhongyuan Wang}, {and} \bibinfo{person}{Ji{-}Rong Wen}.}
  \bibinfo{year}{2020}\natexlab{}.
\newblock \showarticletitle{S3-Rec: Self-Supervised Learning for Sequential
  Recommendation with Mutual Information Maximization}. In
  \bibinfo{booktitle}{\emph{CIKM'20}}. \bibinfo{pages}{1893--1902}.
\newblock


\bibitem[Zhu et~al\mbox{.}(2020)]%
        {conf/ijcai/ZhuWCLZ20}
\bibfield{author}{\bibinfo{person}{Feng Zhu}, \bibinfo{person}{Yan Wang},
  {et~al\mbox{.}}} \bibinfo{year}{2020}\natexlab{}.
\newblock \showarticletitle{A Graphical and Attentional Framework for
  Dual-Target Cross-Domain Recommendation}. In
  \bibinfo{booktitle}{\emph{IJCAI'20}}. \bibinfo{pages}{3001--3008}.
\newblock


\bibitem[Zhu et~al\mbox{.}(2021)]%
        {conf/sigir/ZhuGZXXZL021}
\bibfield{author}{\bibinfo{person}{Yongchun Zhu}, \bibinfo{person}{Kaikai Ge},
  \bibinfo{person}{Fuzhen Zhuang}, \bibinfo{person}{Ruobing Xie},
  \bibinfo{person}{Dongbo Xi}, \bibinfo{person}{Xu Zhang},
  \bibinfo{person}{Leyu Lin}, {and} \bibinfo{person}{Qing He}.}
  \bibinfo{year}{2021}\natexlab{}.
\newblock \showarticletitle{Transfer-Meta Framework for Cross-domain
  Recommendation to Cold-Start Users}. In \bibinfo{booktitle}{\emph{SIGIR'21}}.
  \bibinfo{pages}{1813--1817}.
\newblock


\bibitem[Zhu et~al\mbox{.}(2022)]%
        {conf/wsdm/ZhuTLZXZLH22}
\bibfield{author}{\bibinfo{person}{Yongchun Zhu}, \bibinfo{person}{Zhenwei
  Tang}, \bibinfo{person}{Yudan Liu}, \bibinfo{person}{Fuzhen Zhuang},
  \bibinfo{person}{Ruobing Xie}, \bibinfo{person}{Xu Zhang},
  \bibinfo{person}{Leyu Lin}, {and} \bibinfo{person}{Qing He}.}
  \bibinfo{year}{2022}\natexlab{}.
\newblock \showarticletitle{Personalized Transfer of User Preferences for
  Cross-domain Recommendation}. In \bibinfo{booktitle}{\emph{WSDM'22}}.
  \bibinfo{pages}{1507--1515}.
\newblock


\end{thebibliography}
